# Universal patterns underlying ongoing wars and terrorism

Neil F. Johnson[1,6], Mike Spagat[2,6], Jorge A. Restrepo[3,6], Oscar Becerra[6], Juan Camilo Bohórquez[4], Nicolas Suárez[6], Elvira Maria Restrepo[5,6], and Roberto Zarama[4]

[1] *Department of Physics, University of Oxford, Oxford, U.K.*
[2] *Department of Economics, Royal Holloway College, University of London, Egham, U.K.*
[3] *Department of Economics, Universidad Javeriana, Bogotá, Colombia*
[4] *Department of Industrial Engineering, Universidad de los Andes, Bogotá, Colombia*
[5] *Faculty of Economics, Universidad de Los Andes, Bogotá, Colombia*
[6] *CERAC, Conflict Analysis Resource Center, Bogotá, Colombia*

**ABSTRACT**

**We report a remarkable universality in the patterns of violence arising in three high-profile ongoing wars, and in global terrorism. Our results suggest that these quite different conflict arenas currently feature a common type of enemy, i.e. the various insurgent forces are beginning to operate in a similar way regardless of their underlying ideologies, motivations and the terrain in which they operate. We provide a microscopic theory to explain our main observations. This theory treats the insurgent force as a generic, self-organizing system which is dynamically evolving through the continual coalescence and fragmentation of its constituent groups.**

**NOTE**

*This document provides extensive tests and checks of the findings reported in 2005 in preprint physics/0506213 available at: http://xxx.lanl.gov/abs/physics/0506213*

*The results of these tests and checks confirm all the main findings of preprint physics/0506213.*

*Please see PART 5 of the Appendices (i.e. end of this document) for an overview.*

Emails: M.Spagat@rhul.ac.uk, n.johnson@physics.ox.ac.uk

The recent terrorist attacks in London, Madrid and New York (i.e. '9/11'), killed many people in a very short space of time. At the other extreme, the world currently hosts several longer-term 'local' conflicts within specific countries, in which there is a steady stream of new casualties every day. Examples include the wars in Iraq and Afghanistan, and the longer-term guerrilla war in Colombia which is taking place against a back-drop of drug-trafficking and Mafia activity. The origins, motivations, locations and durations of these various conflicts are very different -- hence one would expect the details of their respective dynamical evolution to also be very different.

Here we report the remarkable finding that identical patterns of violence are currently emerging within these different international arenas. Not only have the wars in Iraq and Colombia evolved to yield a *same* power-law behavior, but this behavior is currently of the *same* quantitative form as the war in Afghanistan and global terrorism in non-G7 countries. Our findings suggest that the dynamical evolution of these various examples of modern conflict has less to do with geography, ideology, ethnicity or religion and much more to do with the day-to-day mechanics of human insurgency; the respective insurgent forces are effectively becoming identical in terms of how they operate. Our findings are backed up by extensive statistical tests on carefully prepared datasets, as discussed in the *Appendices*. We also provide a microscopic mathematical model which describes how such a 'common enemy' might be operating. The model represents the insurgent force as an evolving population of attack units whose destructive potential varies over time. Not only is the model's power-law behavior in excellent agreement with the data from Iraq, Colombia and non-G7 terrorism, it is also consistent with data obtained from the recent war in Afghanistan. These findings suggest that modern insurgent wars tend to be driven by the same underlying mechanism: the continual coalescence and fragmentation of attack units.

Power-law distributions are known to arise in a large number of physical, biological, economic and social systems[1]. In the present context, a power-law distribution means that the probability that an event will occur with $x$ victims is given by $p(x) = Cx^{-\alpha}$ over a reasonably wide range of $x$, with $C$ and $\alpha$ positive coefficients. This in turn implies that a graph of $\log[P(X \geq x)]$ vs. $\log(x)$ will be a straight line with a negative slope of magnitude $\alpha - 1$.[i] Previous studies have shown that the distribution obtained from 'old' wars, 1816-1980, exhibits a power-law with $\alpha = 1.80(9)$[1-4],[ii] However each data-point in these studies represents the total casualty figure for one particular war. By contrast, our analysis looks at the pattern of casualties arising *within a given war*. Casualty numbers in global terrorist events, from 1968 to the present, are also known to obey power laws where in this case each data point is a terrorist attack[5]: $\alpha = 2.5(1)$ for non-G7 countries while $\alpha = 1.71(3)$ for G7 countries[5]. We find that both Iraq and Colombia exhibit power-law behavior and that their power-law coefficients are currently close to 2.5, which is exactly the value characterizing non-G7 terrorism. We also find power-law behavior for Afghanistan with a coefficient near 2.5.

Figure 1 shows log-log plots of the fraction of all recorded events with $x$ or more victims, $P(X \geq x)$, versus $x$. As a result of extensive statistical testing, as discussed in the *Appendices*, we conclude that this time-aggregated data from each war follows a power-law over a wide range of $x$ values.

---

[i] We will refer to $P(X \geq x)$ as the cumulative distribution obtained from $p(x)$.

[ii] Numbers in parentheses give the standard error on the trailing figure in each case.

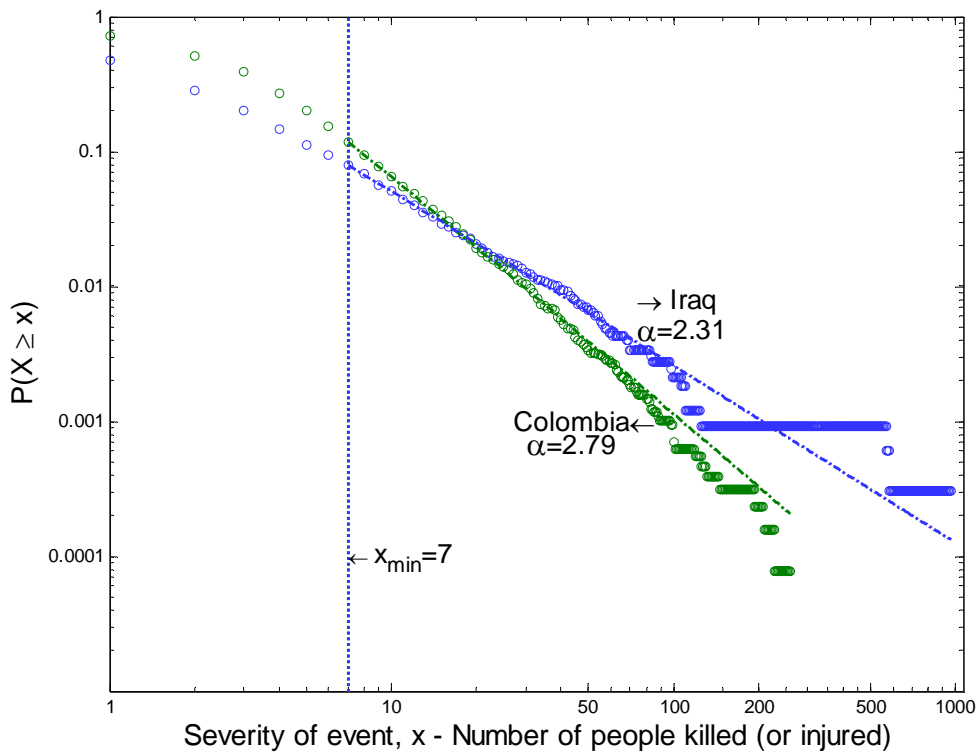

**Figure 1** Log-log plots of the cumulative distributions $P(X \geq x)$ describing the total number of events with severity greater than $x$, for the ongoing wars in Iraq (blue) and Colombia (green). For Iraq, the severity is the lower estimate of deaths from the CERAC Integrated Iraq Dataset covering the period 5/1/03 to 10/23/05. For Colombia, the severity is the total number of deaths plus injuries from the CERAC Colombia Conflict Dataset[6] for 1988 through 2004. Each line indicates the most likely power-law fit for the data above $x_{min}$ (see *Appendices*). The estimated $x_{min}$ values turn out to be the same for both conflicts.

As well as observing power-law behavior for the time-aggregated data shown in Figure 1, we also find power-law behavior over smaller time-windows. We can follow the time-evolution of the power-law coefficient $\alpha$ by sliding this time-window through the data-series. Figure 2 shows the resulting $\alpha$ values as a function of time, in addition to the corresponding 95% confidence bands. Remarkably, these alpha values have recently become very close to 2.5 which also happens to be the value for global terrorism in non-G7 countries and the war in Afghanistan (see below). The implication is that these wars and global non-G7 terrorism are showing very similar underlying behavior.

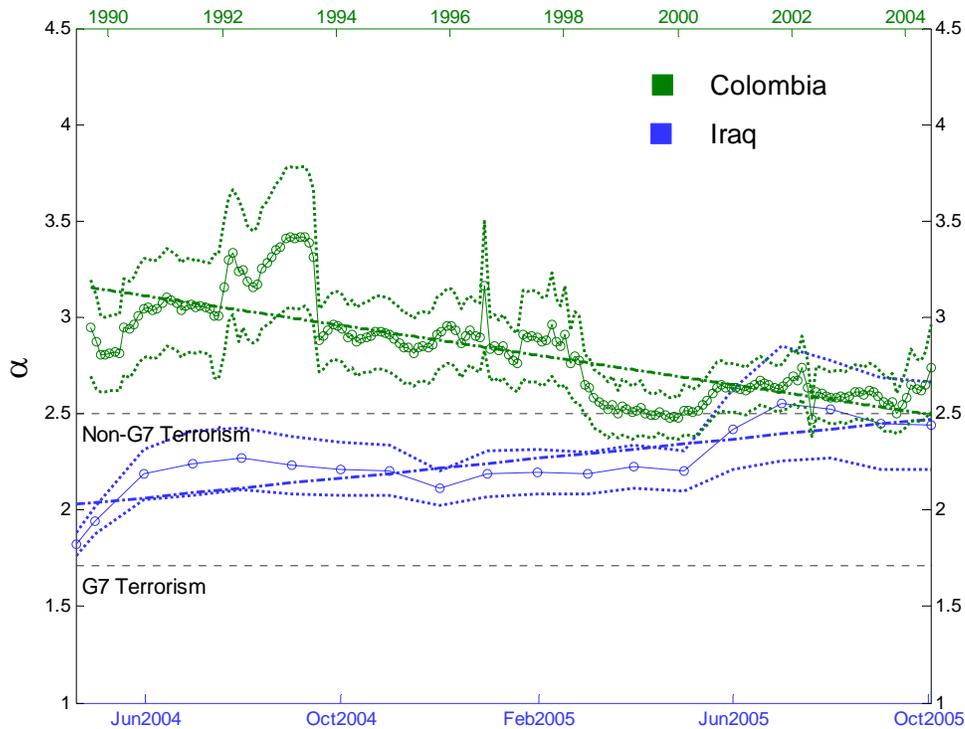

**Figure 2** The variation through time of the power-law coefficient $\alpha$ for Iraq and Colombia, shown respectively as blue and green lines with square symbols together with their corresponding 95% confidence bands. For Iraq and Colombia the time-windows are 400 and 800 days respectively, with each time-window sliding forward one month at a time. The straight lines are fits through these points, and suggest a current value of $\alpha \approx 2.5$ for both wars. The values for G7 and non-G7 terrorism are also shown[5].

But why should a seemingly universal value of 2.5 emerge for Iraq, Colombia and non-G7 global terrorism? Standard physical mechanisms for generating power laws make little sense in the context of Colombia or Iraq[1]. One might initially guess that casualties would arise in rough proportion to the population sizes of the places where insurgent groups attack: given that city populations may follow a power law[1], it is conceivable that this would also produce power laws for the severity of attacks. However, we have tested this hypothesis against our Colombia data and it is resoundingly rejected.

Figure 3 shows a model of modern 'generic' insurgent warfare which we have developed. Full details are given in the *Appendices*. Our model proposes that the insurgent force operates as a dynamically evolving population of fairly self-contained units, which we call 'attack units'. Each attack unit has a particular 'attack strength' characterizing the number of casualties which typically arise in an event involving this attack unit. As time evolves, these attack units either join forces with other attack units (i.e. coalescence) or break up (i.e. fragmentation). Eventually this on-going process of coalescence and fragmentation reaches a dynamical steady-state which is solvable analytically, yielding a power-law with coefficient $\alpha = 2.5$ (see *Appendices*). The combination of these empirical and analytical findings suggests that similar distributions of attack units are emerging in Colombia, Iraq and in non-G7 global terrorism, with each attack unit in an ongoing state of coalescence and fragmentation. Furthermore, our model offers the following interpretation for the evolution of $\alpha$ observed in Figure 2: The Iraq war began as a conventional confrontation between large armies, but continuous pressure applied to the Iraqis by coalition forces has fragmented the insurgency into a structure in which smaller attack units, characteristic of non-G7 global terrorism, now predominate. In Colombia, on the other hand, the guerrillas in the early 1990's had even less ability than global terrorists to coalesce into high-impact units but have gradually been acquiring comparable capabilities.

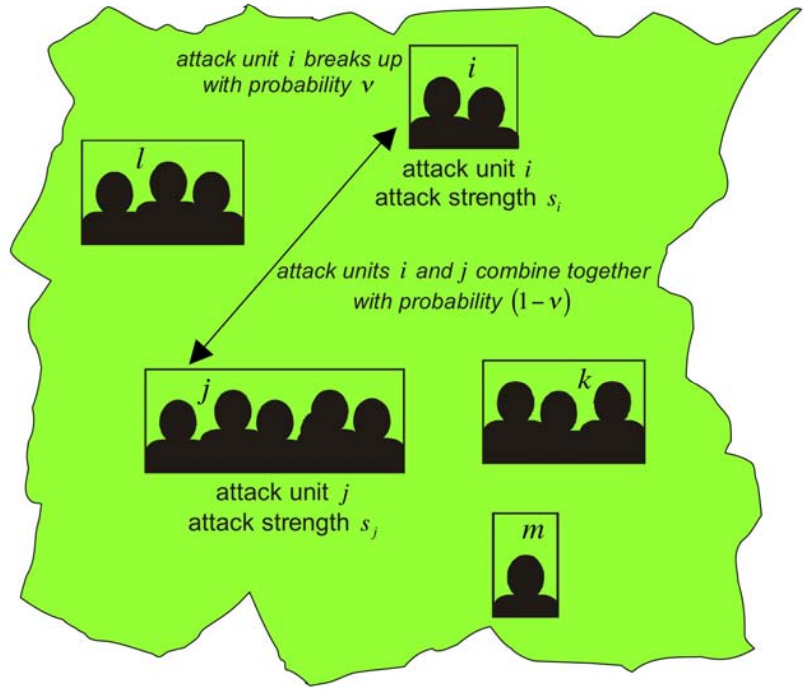

Figure 3

Figure 3   Our analytically-solvable model describing modern insurgent warfare. The insurgent force comprises attack units, each of which has a particular attack strength. As indicated by the dark shadows, this attack strength represents the number of casualties which the attack unit would typically inflict in a conflict event. The total attack strength is being continually re-distributed through coalescence and fragmentation of the attack units.

These empirical and analytical findings lead us to speculate that power-law patterns will emerge within any modern asymmetric war which is being fought by loosely-organized insurgent groups. Such generic behavior can be explained using theoretical models which describe group formation, rather than having to invoke case-specific issues such as politics or geography. Although future wars will provide the ultimate test of such a claim, Figure 4 provides some further supporting evidence. In particular, the casualty figures from the recent Afghanistan war follow a power-law with $\alpha$ very close to 2.5. Additional supporting cases will be presented elsewhere.

Finally, we note that we have obtained access to preliminary databases from a range of other insurgent-like wars in recent history. Since we have not been able to subject their event entries to the same rigorous scrutiny as those for Iraq, Colombia and Afghanistan, any estimates of power-laws from them would be subject to very large error-bars. However, it is interesting to note that if we carry out the same exercise as in this paper, and treat the resulting estimates at face value, then the average value of $\alpha$ obtained across a broad platform of such wars –including Casamance (Senegal), Indonesia, Israel and Northern Ireland – is between 2.4 and 2.5. A detailed discussion of these other conflicts will be given elsewhere after we have been able to cross-check the accuracy of the respective databases.

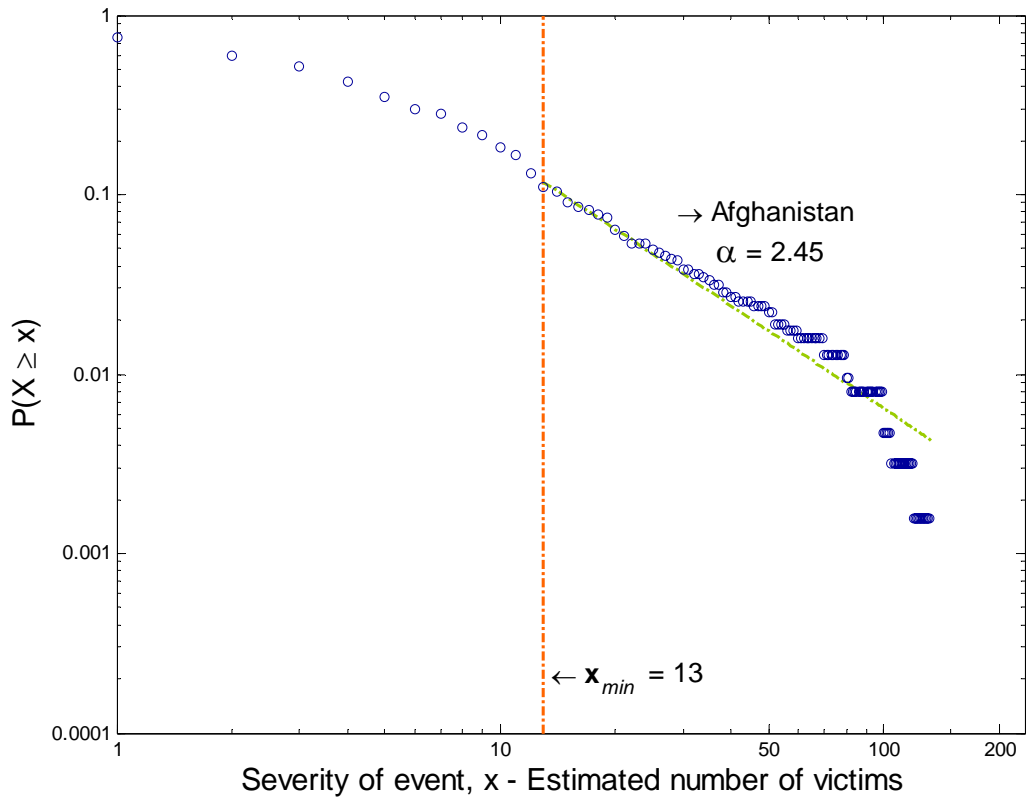

**Figure 4** Log-log plots of the cumulative distribution $P(X \geq x)$ describing the total number of events with killings greater than $x$ for Afghanistan. The data is the lower estimate from the CERAC Integrated Afghanistan Dataset, covering the period 9/9/01 to 7/31/04. The power-law fit, together with the estimated $\alpha$ and $x_{\min}$, are shown.

**Acknowledgements:** We thank Wilmer Marín and Simón Mesa of CERAC for research assistance. The Department of Economics of Royal Holloway College provided funds to build the CERAC Colombia Conflict Database. J.A. Restrepo acknowledges funding from Banco de la República, Colombia. J.C. Bohórquez acknowledges funding from the Department of Industrial Engineering of Universidad de los Andes. N.F. Johnson thanks the Zilkha Fund (Lincoln College, U.K.) for funding a research trip to Colombia.

# **Appendices**

**PART 1**:  Details of the model mentioned in the paper

**PART 2**:  Data and methods

**PART 3**:  Table and figures confirming the robustness of our results

**PART 4**:  Events excluded from the IBC dataset.

**PART 5**: An explanation of how the results in this document provide an extensive test, check and confirmation of the findings reported earlier in preprint physics/0506213 which is available at:

http://xxx.lanl.gov/abs/physics/0506213

## **PART 1: Details of the model mentioned in the paper**

Here we provide details of our model of modern insurgent warfare, which we introduced in the main paper. Our goal is to provide a plausible model to explain (i) why power-law behavior is observed in the Colombia, Iraq and Afghanistan wars, and in non-G7 global terrorism, and (ii) why the power-law coefficients for the Colombia and Iraq wars should both be currently around the value of 2.5, observed in non-G7 terrorism and Afghanistan.

Our model bears some similarity to a model of herding by Cont and Bouchaud[iii], and is a direct adaptation of the Eguiluz-Zimmerman model of herding in financial markets[iv]. The analytical derivation which we present, is an adaptation of earlier formalism laid out by D'Hulst and Rodgers[v], and also draws heavily on the material in the book *Financial Market Complexity* by Neil F. Johnson, Paul Jefferies and Pak Ming Hui (Oxford University Press, 2003). One of us (NFJ) is extremely grateful to Pak Ming Hui for detailed correspondence about the Eguiluz-Zimmerman model of financial markets, the associated formalism, and its extensions – and also for discussions involving the present model.

As suggested by Figure 3 in the paper, our model is based on the plausible notion that the total attack capability of a modern insurgent force is being continually re-distributed. Based on our intuition about such 'new' wars, we consider the insurgent force to be made

---

[iv] R. Cont and J.P. Bouchaud, Macroeconomic Dynamics **4**, 170 (2000)

[iv] V.M. Eguiluz and M.G. Zimmerman, Phys. Rev. Lett. **85**, 5659 (2000)

[v] R. D'Hulst and G.J. Rodgers, Eur. Phys. J. B **20**, 619 (2001). See also Y. Xie, B.H. Wang, H. Quan, W. Yang and P.M. Hui, Phys. Rev. E **65**, 046130 (2002).

up of ***attack units***, with each one having a certain ***attack strength*** (see below for a detailed discussion). It is reasonable to expect that the total attack strength for the entire insurgent force would change fairly slowly over time. At any particular instant, this total attack strength is distributed (i.e. partitioned) among the various attack units -- moreover the composition of these attack units, and hence their relative attack strengths, will evolve in time as a result of an on-going process of coalescence (i.e. combination of attack units) and fragmentation (i.e. breaking up of attack units). Such a process of coalescence and fragmentation is realistic for an insurgent force in a guerilla-like war, and will be driven by a combination of planned decisions and opportunistic actions by both the insurgent force and the incumbent force. For example, separate attack units might coalesce prior to an attack, or an individual attack unit might fragment in response to a crackdown by the incumbent force. Here we will model this process of coalescence and fragmentation as a stochastic process.

Each attack unit carries a specific label $i, j, k, \ldots$ and has an attack strength denoted by $s_i, s_j, s_k, \ldots$ respectively. By ***attack unit*** we have in mind a group of people, weapons, explosives, machines, or even information, which organizes itself to act as a single unit. In the case of people, this means that they are probably connected by location (e.g. they are physically together) or connected by some communication system. In the case of a piece of equipment, this means that it is readily available for use by members of a particular group. The simplest scenario is to just consider people, and in particular a group of insurgents which are in such frequent contact that they are able to act as a single group. However, we emphasize that an attack unit may also consist of a combination of people and objects – for example, explosives plus a few people, such as in the case of suicide bombers. Such an attack unit, while only containing a few people, could have a high attack strength. In addition, information could also be a valuable part of an attack unit. For example, a lone suicide bomber who knows when a certain place will be densely populated (e.g. a military canteen at lunchtimes) and who knows how to get into such a place unnoticed, will also represent an attack unit with a high attack strength. We define the **attack strength,** $s_i$, of a given attack unit $i$, as the average number of people who are typically injured or killed as the result of an event involving attack unit $i$. In other words, a typical event (e.g. attack or clash) involving group $i$ will lead to the injury or death of $s_i$ people. This definition covers both the case of one-sided attacks by attack unit $i$ (since in this case, all casualties are due to the presence of attack unit $i$) and it also covers two-sided clashes (since presumably there would have been no clash, and hence no casualties, if unit $i$ had not been present).

We take the sum of the attack strengths over all the attack units (i.e. the total attack strength of the insurgent force) to be equal to $N$. From the definition of attack strength, it follows that $N$ represents the maximum number of people which would be injured or killed in an event, on average, if the entire insurgent force were to act together as a single attack unit. Mathematically, $\sum_{i,j,k,...} s_i = N$. For any significant insurgent force, one would expect $N >> 1$, however the power-law results that we will derive do not depend on any particular choice of $N$. In particular, the power-law result derived in this section concerning the average number $n_s$ of attack units having a given attack strength $s$, is invariant under a global magnification of scale (as are all power-laws).

Hence our model becomes, in mathematical terms, one in which the total attack strength $N$ is being continually re-distributed among attack groups as a result of an ongoing process of coalescence and fragmentation. As a further clarification of our terminology, we will now discuss the two limiting cases which, for convenience, we classify as the 'complete coalescence' and 'complete fragmentation' limits:

- 'Complete coalescence' limit: Suppose the conflict is such that all the attack units join together or *coalesce* into a single large attack unit. This would correspond to amassing all the available combatants and weaponry in a single place – very much like the armies of the past would amass their entire force on the field of battle. Hence there is one large attack unit, which we label as $i$ and which has an attack strength $N$. All other attack units disappear. Hence $s_i \to N$. This complete coalescence limit has the *minimum* possible number of attack units (i.e. one) but the *maximum* possible attack strength (i.e. $N$) in that attack unit.

- 'Complete fragmentation' limit: Suppose the conflict is such that all the attack units *fragment* into ever smaller attack units. Eventually all attack units will have attack strength equal to one. Hence $s_i \to 1$ for all $i = 1, 2, \ldots, N$. This would correspond to all combatants operating essentially individually. This complete fragmentation limit has the *maximum* possible number of attack units (i.e. $N$) but the *minimum* possible attack strength per attack unit (i.e. one).

In practice, of course, one would expect the situation to lie between these two limits of complete coalescence and complete fragmentation. In particular, it seems reasonable to expect that the attack units and their respective attack strengths will evolve in time within a given war. Indeed, one can envisage that the attack units will occasionally either break up into smaller groups (i.e. smaller attack units) or join together to form larger ones. The reasons are plentiful why this should occur: for example, the opposing forces (e.g. the Colombian Armed Forces and National Police in Colombia or Coalition Forces in Iraq or Afghanistan) may be applying pressure in terms of searching for hidden insurgent groups.

Hence these insurgent groups (i.e. attack units) might either decide, or be forced, to break up in order to move more quickly, or in order to lose themselves in the towns or countryside.

Hence attack units with different attack strengths will continually mutate via coalescence and fragmentation yielding a 'soup' of attack units with a range of attack strengths. At any one moment in time, this 'soup' corresponds mathematically to partitioning the total $N$ units of attack strength which the insurgent army possesses. The analysis which we now present suggests that the current states of the guerilla/insurgency wars in Colombia, Iraq and Afghanistan all correspond to the steady-state limit of such an on-going coalescence-fragmentation process. It also suggests that such a process might also underpin the acts of terrorism in non-G7 countries, and that such terrorism is characteristic of some longer-term 'global war'.

Against the backdrop of on-going fragmentation and coalescence of attack units, we suppose that each attack unit has a given probability $p$ of being involved in an event in a given time-interval, regardless of its attack strength. For example, $p$ could represent the probability that an arbitrarily chosen attack unit comes across an undefended target – or vice versa, the probability that an arbitrarily chosen attack unit finds itself under attack. In these instances, $p$ should be relatively insensitive to the actual attack strength of the attack unit involved: given this condition, the results which we shall derive for the distribution of attack strengths can also be used to describe the distribution of events having a given severity. When obtaining our results, we shall assume that the war has been underway for a long time and hence some kind of steady-state has been reached. This latter assumption is plausible for the conflicts in Colombia, Iraq and Afghanistan, and also for non-G7 terrorism. Given these considerations, it follows that if there are on average $n_s$ attack units of a given attack strength $s$, then the number of events involving an attack unit of attack strength $s$ will be proportional to $n_s$. We assume, quite realistically, that only one insurgent attack group participates in a given event. For example, attacks in which 10 people are killed are due to an attack by a unit of attack strength 10 as opposed to simultaneous attacks by a unit of strength 6 and a unit of strength 4 (i.e. 6+4=10). Hence the number of events in which $s$ people were killed and/or injured, is just proportional to $n_s$. In other words, the histogram, and hence power-law, that we will derive for the dependence of $n_s$ on $s$, will also describe the number of events with $s$ casualties versus $s$. Indeed, if we consider that an event will typically have duration of $T$, and that there will only be a few such events in a given interval $T$, then these results should also appear similar to the distribution describing the number of intervals of duration $T$ in which there were $s$ casualties, versus $s$. This is indeed what we have found in our analysis of the empirical data.

Given the above discussion, our task of analyzing and deducing the average number of events with $s$ casualties versus $s$ over a given period of time becomes equivalent to the task of analyzing and deducing the average number $n_s$ of attack units of a given attack strength $s$ in that same period of time. This is what we will now calculate. We will start by considering a reasonable mechanism for the coalescence and fragmentation of attack groups, before then finally deducing analytically the corresponding power-law behavior and obtaining a power-law coefficient equal to 2.5.

Consider an arbitrary attack unit $i$ with attack strength $s_i$. At any one instant in time, labeled $t$, we assume that this attack unit will do any one of three things:

    a) fragment (i.e. break up) into $s_i$ attack units of attack strength equal to 1. This feature aims to mimic an insurgent group which decides, either voluntarily or

involuntarily, to split itself up (e.g. in order to reduce the chance of being captured and/or to mislead the enemy).

    b) coalesce (i.e. combine) with another attack unit $j$ of attack strength $s_j$, hence forming a single attack unit of attack strength $s_i + s_j$. This feature mimics two insurgent groups finding each other by chance (e.g. in the Colombian jungle) or deciding via radio communication to meet up and join forces.

    c) remain unchanged. This feature mimics an insurgent group which has no reason to split itself up, nor does it have a reason to coalesce with another attack unit.

To implement this process for each possible attack unit and at every timestep, we proceed as follows. At each timestep, we choose an attack unit $i$ at random but with a probability which is proportional to its attack strength $s_i$. With a probability $\nu$, this attack unit $i$ with attack strength $s_i$ *fragments* into $s_i$ attack units with attack strength 1. A justification for choosing attack unit $i$ with a probability which is proportional to its attack strength is as follows: attack units with higher attack strength are likely to be bigger and hence will either run across the enemy more and/or be more actively sought by the enemy. By contrast, with a probability $(1-\nu)$, the chosen attack unit $i$ instead *coalesces* with another attack unit $j$ which is chosen at random, but again with a probability which is proportional to its attack strength $s_j$. The two attack units of attack strengths $s_i$ and $s_j$ then combine to form a bigger attack unit of attack strength $s_i + s_j$. The justification for choosing attack unit $j$ for coalescence with a probability which is proportional to its attack strength is as follows: it is presumably risky to combine attack units, since it must involve at least one message passing between the two units in order to coordinate their actions. Hence it becomes increasingly less worthwhile to combine attack units as the attack units get smaller.

This model is thus characterized by a single parameter $\nu$. The connectivity among the attack units is driven by the dynamics of the model. For very small $\nu$ (i.e. much less than 1), the attack units steadily coalesce. This leads to the formation of large attack units. In the other limit of $\nu \rightarrow 1$, the system consists of many attack units with attack strength close to 1. A value of $\nu = 0.01$ corresponds to about one fragmentation in every 100 iterations. In what follows, we assume that $\nu$ is small since the process of fragmentation should not be very frequent for any insurgent force which is managing to sustain an ongoing war. Indeed if such fragmentation were very frequent, then this would imply that the insurgents were being so pressured by the incumbent force that they had to fragment at nearly every timestep. Hence that particular war should not last very long. It turns out that infrequent fragmentations are sufficient to yield a steady-state process, and will also yield the power-law behavior which we observe for Colombia and Iraq.

A typical result obtained from numerical simulations of this model, for the distribution of $n_s$ versus attack strength $s$ in the long-time limit (i.e. steady-state), is shown below in terms of $n_s/n_1$:

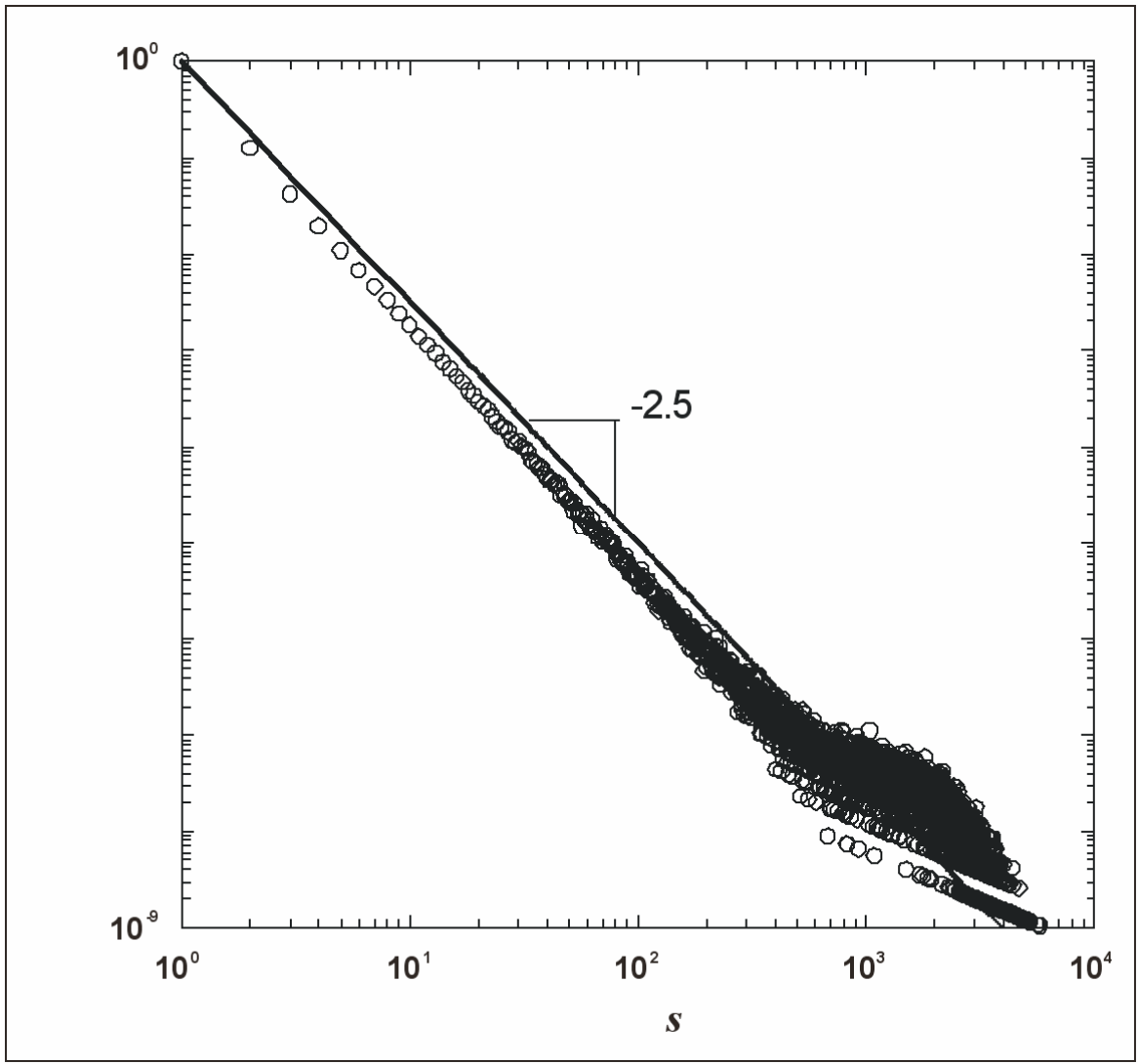

Figure S1: Log-log plot of the number of attack units with attack strength *s*, versus attack strength *s*. Here $N = 10,000$ and $\nu = 0.01$. The results are obtained from a numerical simulation of the model. The initial conditions of this numerical simulation are such that all attack units have size 1 -- however, changing this initial condition does not change the main results which emerge in the steady state. As time evolves, the attack units undergo coalescence and fragmentation as described in the text. In the long-time limit, the system reaches a steady-state with power-law dependence as shown in the figure and with an associated power-law

coefficient of 2.5 (i.e. 5/2). The deviation from power-law behavior at large *s* is simply due to the finite value of *N*: since there can be no attack unit with attack strength greater than *N*, the finite size of *N* distorts the power-law as *s* approaches *N*.

We now provide an *analytic* derivation of the observed power-law behavior, and specifically the power-law coefficient 2.5, in the steady-state (i.e. long-time) limit.

One can write a dynamical equation for the evolution of the model at different levels of approximation. For example, one could start with a microscopic description of the system by noting that at any moment in time, the entire insurgent army can be described by a partition $\{l_1, l_2, \ldots, l_N\}$ of the total attack strength *N* into *N* attack units. Here $l_s$ is the number of attack units of attack strength *s*. For example $\{0,0,\ldots,1\}$ corresponds to the 'complete coalescence' limit discussed earlier, in which all the attack strength is concentrated in one big attack unit. By contrast, $\{N,0,\ldots,0\}$ corresponds to the 'complete fragmentation' limit in which all the attack units have attack strength of 1 (i.e. there are *N* attack units of attack strength 1). Clearly, the total amount of attack strength is conserved $\sum_{i=1}^{N} i l_i = N$. All that happens is that the way in which this total attack strength *N* is *partitioned* will change in time.

In principle, the dynamics could be described by the time-evolution of the probability function $p[l_1, l_2, \ldots, l_N]$: in particular, taking the continuous-time limit would yield an equation for $dp[l_1, l_2, \ldots, l_N]/dt$ in terms of transitions between partitions. For example, the fragmentation of an attack unit of attack strength *s* leads to a transition from the partition $\{l_1, \ldots, l_s, \ldots, l_N\}$ to the partition $\{l_1 + s, \ldots, l_s - 1, \ldots, l_N\}$. For our purposes, however, it is more convenient to work with the *average* number $n_s$ of attack units of attack strength *s*, which can be written as $n_s = \sum_{\{l_1,...,l_N\}} p[l_1,...,l_s,...,l_N] \cdot l_s$. The sum is over all possible partitions.

Since $p[l_1,...,l_N]$ evolves in time, so does $n_s[t]$. After the transients have died away, the system is expected to reach a steady-state in which $p[l_1,...,l_N]$ and $n_s[t]$ become time-independent. The time-evolution of $n_s[t]$ can be written down either by intuition, or by invoking a mean-field approximation to the equation for $dp[l_1,l_2,...,l_N]/dt$. Taking the intuitive route, one can immediately write down the following dynamical equations in the continuous-time limit:

$$\frac{\partial n_s}{\partial t} = -\frac{\nu s n_s}{N} + \frac{(1-\nu)}{N^2}\sum_{s'=1}^{s-1} s' n_{s'}(s-s')n_{s-s'} - \frac{2(1-\nu)s n_s}{N^2}\sum_{s'=1}^{\infty} s' n_{s'} \quad \text{for } s \geq 2 \quad (0.1)$$

$$\frac{\partial n_1}{\partial t} = \frac{\nu}{N}\sum_{s'=2}^{\infty}(s')^2 n_{s'} - \frac{2(1-\nu)n_1}{N^2}\sum_{s'=1}^{\infty} s' n_{s'} \quad (0.2)$$

The terms on the right-hand side of Equation (0.1) represent all the ways in which $n_s$ can change. The first term represents a decrease in $n_s$ due to the fragmentation of an attack unit of attack strength $s$: this happens only if an attack unit of attack strength $s$ is chosen and if fragmentation then follows. The former occurs with probability $s n_s/N$ (see earlier discussion) and the latter with probability $\nu$. The second term represents an increase in $n_s$ as a result of the merging of an attack unit of attack strength $s'$ with an attack unit of attack strength $(s-s')$. The third term describes the decrease in $n_s$ due to the merging of an attack unit of attack strength $s$ with any other attack unit. For the $s=1$ case described by Equation (0.2), the chosen attack unit remains isolated; thus Equation (0.2) does not have a contribution like the first term of Equation (0.1). The first term which appears in Equation (0.2) reflects the increase in the number of attack units of attack strength equal to 1, due to fragmentation of an attack unit. Similarly to Equation (0.1), the last term of Equation (0.2) describes the merging of an attack unit with attack strength 1, with an attack unit of any other attack strength. Equations (0.1) and (0.2) are so-called 'master equations' describing the dynamics within the model. Note that for simplicity, we are only considering fragmentation into attack units of attack strength 1. However this could be generalized – indeed, we will look at more general fragmentations in future publications.

In the long-time steady state limit, Equations (0.1) and (0.2) yield:

$$s n_s = \frac{(1-v)}{(2-v)N} \sum_{s'=1}^{s-1} s' n_{s'} (s-s') n_{s-s'} \quad \text{for } s \geq 2 \qquad (0.3)$$

$$n_1 = \frac{v}{2(1-v)} \sum_{s'=2}^{\infty} (s')^2 n_{s'} \qquad (0.4)$$

Equations of this type are most conveniently treated using the general technique of 'generating functions'. As the name suggests, these are functions which can be used to generate a range of useful quantities. Consider

$$G[y] = \sum_{s'=0}^{\infty} s' n_{s'} y^{s'} \qquad (0.5)$$

where $y = e^{-\omega}$ is a parameter. Note that $s n_s / N$ is the probability of finding an attack unit of attack strength $s$. If $G[y]$ is known, $s n_s$ is then formally given by

$$s n_s = \frac{1}{s!} G^{(s)}[0] \qquad (0.6)$$

where $G^{(s)}[y]$ is the $s$-th derivative of $G[y]$ with respect to $y$. $G^{(s)}[y]$ can be decomposed as

$$G[y] = n_1 y + \sum_{s'=2}^{\infty} s' n_{s'} y^{s'} \equiv n_1 y + g[y] \qquad (0.7)$$

where the function $g[y]$ governs the attack-units' attack-strength distribution $n_s$ for $s \geq 2$. The next task is to obtain an equation for $g[y]$. This can be done in two ways. One could either write down the terms in $(g[y])^2$ explicitly and then make use of Equation (0.3), or one could construct $g[y]$ by multiplying Equation (0.3) by $e^{-\omega s}$ and then summing over $s$. The resulting equation is:

$$(g[y])^2 - \left(\frac{2-\nu}{1-\nu}N - 2n_1 y\right)g[y] + n_1^2 y^2 = 0 \tag{0.8}$$

First we solve for $n_1$. From Equation (0.7), $g[1] = G[1] - n_1 = N - n_1$. Substituting $n_1 = N - g[1]$ into Equation (0.8) and setting $y = 1$, yields

$$g[1] = \frac{1-\nu}{2-\nu}N \tag{0.9}$$

Hence
$$n_1 = N - g[1] = \frac{1}{2-\nu}N \tag{0.10}$$

To obtain $n_s$ with $s \geq 2$, we need to solve for $g[y]$. Substituting Equation (0.10) for $n_1$, Equation (0.8) becomes

$$(g[y])^2 - \left(\frac{2-\nu}{1-\nu}N - \frac{2N}{2-\nu}y\right)g[y] + \frac{N^2}{(2-\nu)^2}y^2 = 0 \tag{0.11}$$

Equation (0.11) is a quadratic equation for $g[y]$ which can be solved to obtain

$$g[y] = \frac{(2-\nu)N}{4(1-\nu)}\left(1 - \sqrt{1 - \frac{4(1-\nu)}{(2-\nu)^2}y}\right)^2$$
$$= \frac{(2-\nu)N}{4(1-\nu)}\left(2 - \frac{4(1-\nu)}{(2-\nu)^2}y - 2\sqrt{1 - \frac{4(1-\nu)}{(2-\nu)^2}y}\right). \tag{0.12}$$

Using the expansion[vi]

$$(1-x)^{1/2} = 1 - \frac{1}{2}x - \sum_{k=2}^{\infty}\frac{(2k-3)!!}{(2k)!!}x^k, \tag{0.13}$$

we have

---

[vi] The 'double factorial' operator !! denotes the product: $n!! = n(n-2)(n-4)\ldots$

$$g[y] = \frac{(2-\nu)N}{2(1-\nu)} \sum_{k=2}^{\infty} \frac{(2k-3)!!}{(2k)!!} \left( \frac{4(1-\nu)}{(2-\nu)^2} y \right)^k. \tag{0.14}$$

Comparing the coefficients in Equation (0.14) with the definition of $g[y]$ in Equation (0.7), the probability of finding an attack unit of attack strength $s$ is given by:

$$\frac{s n_s}{N} = \frac{(2-\nu)}{2(1-\nu)} \frac{(2s-3)!!}{(2s)!!} \left( \frac{4(1-\nu)}{(2-\nu)^2} \right)^s. \tag{0.15}$$

It hence follows that the average number of attack units of attack strength $s$ is

$$\begin{aligned} n_s &= \frac{(2-\nu)}{2(1-\nu)} \frac{(2s-3)!!}{s(2s)!!} \left( \frac{4(1-\nu)}{(2-\nu)^2} \right)^s N \\ &= \frac{(1-\nu)^{s-1}(2s-2)!}{(2-\nu)^{2s-1}(s!)^2} N \end{aligned} \tag{0.16}$$

The $s$-dependence of $n_s$ is implicit in Equation (0.16), with the dominant dependence arising from the factorials. Recall Stirling's series for $\ln[s!]$:

$$\ln[s!] = \frac{1}{2}\ln[2\pi] + \left(s + \frac{1}{2}\right)\ln[s] - s + \frac{1}{12s} - \cdots. \tag{0.17}$$

Retaining the few terms shown in Equation (0.17) is in fact a very good approximation, giving an error of $< 0.05\%$ for $s \geq 2$. This motivates us to take the logarithm of both sides of Equation (0.16) and then apply Stirling's formula to each log-factorial term, as in Equation (0.17). We follow these mathematical steps (which were derived in the M.Phil. thesis of Larry Yip, Chinese University of Hong Kong, who was supervised by Prof. Pak Ming Hui). We hence obtain

$$\begin{aligned} \ln(n_s) &\approx \ln\left( \frac{(1-\nu)^{s-1}}{(2-\nu)^{2s-1}} N \right) + \left(2s - \frac{3}{2}\right)\ln(2s-2) + \ln(e^2) - \frac{1}{2}\ln(2\pi) - (2s+1)\ln(s) \\ &\approx \ln\left( \frac{e^2 4^s (1-\nu)^{s-1}}{2^{\frac{3}{2}}\sqrt{2\pi}(2-\nu)^{2s-1}} N \right) + \left(2s - \frac{3}{2}\right)\ln(s) - \left(3s - \frac{3}{2}\right)\frac{1}{s} - (2s+1)\ln(s) \end{aligned}$$

Combining the terms on the right-hand side into a single logarithm, it follows that

$$n_s \approx \left(\frac{(2-v)e^2}{2^{3/2}\sqrt{2\pi}(1-v)}\right)\left(\frac{4(1-v)}{(2-v)^2}\right)^s \cdot \frac{(s-1)^{2s-3/2}}{s^{2s+1}} N. \qquad (0.18)$$

The $s$-dependence at large $s$ can then be deduced from Equation (0.18):

$$n_s \sim N\left(\frac{4(1-v)}{(2-v)^2}\right)^s s^{-5/2} . \qquad (0.19)$$

The above equation (0.19) can be re-written as follows:

$$n_s \sim N \exp\left(-s \ln\left[\frac{(1-v/2)^2}{(1-v)}\right]\right) s^{-5/2}$$

which shows that at large $s$, there will be an exponential cut-off. This makes sense since the coalescence process for attack units with very large attack strength will always be hampered by the fact that the total insurgent attack strength $N$ is itself finite -- indeed, such a cut-off at large $s$ is also observed in the empirical war data. For sufficiently small values of $v$, the dominant dependence on $s$ over a wide range of intermediate $s$-values will be

$$n_s \sim s^{-5/2} \quad \text{hence} \quad n_s \sim s^{-2.5} \qquad (0.20)$$

This analysis therefore shows analytically that the distribution of attack strengths should follow a power-law with exponent $\alpha = 2.5$ (i.e. 5/2) over a fairly wide range of values of $s$. As discussed earlier, we assume that any particular attack unit could be involved in an event in a given time interval, with a probability $p$ which is independent of its attack size. Hence these power-law results which we have derived for the distribution of attack strengths, will also apply to the distribution of attacks of severity $x$. (Recall that the attack strength $s$ is a measure of the number of casualties in a typical event, and that the severity $x$ of an event is measured as the number of casualties). In other words, the same power-law exponent $\alpha = 2.5$ derived in Equation (0.20), will *also* apply to the distribution of attacks having severity *x*.

**Hence our model predicts that any guerrilla-like war which is characterized by an ongoing process of coalescence and fragmentation of attack units, and hence an ongoing re-distribution of the total attack strength, will have the following properties:**

(i) **The distribution of events with severity *x* will follow a power-law. This finding is consistent with the behavior observed for the aggregated data in the Iraq, Afghanistan and Colombia wars (see Figure 1 of the paper).**

(ii) **The power-law distribution will, in the steady-state (i.e. long-time) limit, have an exponent $\alpha = 2.5$. This is precisely the value which the on-going wars in Colombia and Iraq seem to have reached in recent months (see Figure 2 of the paper) and that Afghanistan exhibits for the whole sample period (see Figure 4 of the paper).**

We will now attempt to go one step further by providing a generalization of the above model in order to offer an explanation for the temporal evolution of the power-law coefficient $\alpha$ which was observed in Figure 2. For technical reasons related to the mathematics of the generating-function approach which we employ, the analytic results which we will obtain for the power-law coefficient $\alpha$ will be more approximate than in the earlier case. However, we have performed numerical simulations to check that these analytic results are still in fact reliable -- furthermore, these analytic results will end up offering important insights into the temporal behavior of $\alpha$ which is observed in Figure 2.

Our generalized model is as follows. Exactly as before, at each timestep an attack unit is randomly picked to be a candidate for the fragmentation-coalescence process. Also as before, the probability of being picked is proportional to its attack strength. We retain this feature of having the probability of being picked as being proportional to the attack strength, since it makes sense to us -- for example, the larger an attack unit, the more likely it will be that it is discovered by opposing forces (and hence may need to decide whether to fragment) or that it itself comes across another attack unit and therefore may need to decide whether to coalesce. In short, it makes sense to us that, on a daily basis, the larger the attack unit is in terms of its attack strength then the more likely it is that it will become a candidate for fragmentation or coalescence. In other words, it seems reasonable to us that the larger the attack strength, the more likely it is that one of the attack unit's constituent parts will become faced with a situation which could lead to fragmentation or coalescence. We now move on to the details of the fragmentation-coalescence process, since this is where the generalization will occur. Suppose a particular attack unit $i$ has been picked at a particular timestep, and that it has an attack strength $s_i$. As before, it is then *selected* for fragmentation with a probability $v$, and for coalescence with a probability $1-v$. However unlike the earlier version, it does not *necessarily* undergo either (hence our use of the word '*selected*'). Instead, the rules are as follows. Suppose it gets selected for fragmentation: it will now fragment with a probability $f[s_i]$ which depends on $s_i$. Suppose instead that it gets selected for coalescence: as before, a second attack unit $j$ is then picked randomly with a probability which is proportional to its attack strength $s_j$. However unlike the earlier version, these two attack units will now coalesce with a probability $f[s_i]f[s_j]$ which depends on $s_i$ and $s_j$. Hence even if a given attack unit is selected for fragmentation or coalescence, neither process is now guaranteed to occur.

The upshot of this generalization is that there are now effectively *three* probabilistic processes at each timestep:

(1) an initial random picking (biased according to attack strength) in order to select an attack unit as a candidate for the fragmentation-coalescence process;

(2) a coin-toss (biased according to the value of $\nu$) in order to decide whether this particular attack unit is being selected for fragmentation or for coalescence. In terms of the mechanics of a war, $\nu$ is the probability that this attack unit is confronted with a situation which might lead to fragmentation, while $1-\nu$ is the probability that this attack unit is confronted with a situation which might lead to coalescence;

(3) another coin-toss (biased according to $f[s_i]$ for fragmentation, and $f[s_i]f[s_j]$ for coalescence) to decide whether the attack unit actually goes through with the fragmentation or coalescence process for which it has been selected. Note that the basic version of the model, which we studied in detail above, corresponds to the simple case with $f[s_i]=1$ and $f[s_j]=1$ for all values of $s_i$ and $s_j$.

In what follows, we demonstrate the specific case where $f[s] \sim s^{-\delta}$ over a reasonably wide range of *s*. In other words, over a reasonably wide range of *s* the probability function $f[s]$ decreases with increasing *s* if $\delta$ is positive, and increases with increasing *s* if $\delta$ is negative. We will also comment on the interpretation and consequences of such a probability function in the setting of the real-world conflicts that we are focusing on.

Since $f[s]$ is a probability, it must of course be normalized and hence cannot strictly take on the form $f[s] \sim s^{-\delta}$ for all *s* and all $\delta$ -- however this does not prevent it from following the functional form $f[s] \sim s^{-\delta}$ over a reasonably wide range of $s$. Analytically, the master equations for $f[s] \sim s^{-\delta}$ can then be readily written down:

$$\frac{\partial n_s}{\partial t} = -\frac{\nu s^{1-\delta} n_s}{N} + \frac{(1-\nu)}{N^2} \sum_{s'=1}^{s-1} (s')^{1-\delta} n_{s'} (s-s')^{1-\delta} n_{s-s'} - \frac{2(1-\nu)s^{1-\delta} n_s}{N^2} \sum_{s'=1}^{\infty} (s')^{1-\delta} n_{s'} \text{ for } s \geq 2 \quad (0.21)$$

$$\frac{\partial n_1}{\partial t} = \frac{\nu}{N} \sum_{s'=2}^{\infty} (s')^{2-\delta} n_{s'} - \frac{2(1-\nu)n_1}{N^2} \sum_{s'=1}^{\infty} (s')^{1-\delta} n_{s'} \quad (0.22)$$

with the meaning of each term being similar to that for Equations (0.1) and (0.2). The steady state equations become

$$s^{1-\delta} n_s = A \sum_{s'=1}^{s-1} (s')^{1-\delta} n_{s'} (s-s')^{1-\delta} n_{s-s'} \quad (0.23)$$

$$n_1 = B \sum_{s'=2}^{\infty} (s')^{2-\delta} n_{s'} \quad (0.24)$$

The constant coefficients $A$ and $B$ are given by

$$A = \frac{1-\nu}{N\nu + 2(1-\nu)\sum_{s'=1}^{\infty}(s')^{1-\delta} n_{s'}} \quad \text{and} \quad B = \frac{N\nu}{2(1-\nu)\sum_{s'=1}^{\infty}(s')^{1-\delta} n_{s'}}$$

Setting $\delta = 0$ in Equations (0.23) and (0.24) recovers Equations (0.3) and (0.4) for the original model. A generating function

$$G[y] = \sum_{s'=0}^{\infty} (s')^{1-\delta} n_{s'} y^{s'} = n_1 y + g[y] \quad (0.25)$$

can be introduced where $g[y] = \sum_{s'=2}^{\infty} (s')^{1-\delta} n_{s'} y^{s'}$ and $y = e^{-\omega}$. The function $g[y]$ satisfies a quadratic equation of the form

$$(g[y])^2 - \left(\frac{1}{A} - 2n_1 y\right) g[y] + n_1^2 y^2 = 0 \quad (0.26)$$

which is a generalization of Equation (0.8). Using $n_1 + g[1] = \sum_{s'=1}^{\infty} (s')^{1-\delta} n_{s'}$ and Equation (0.26), $n_1$ can be obtained as

$$n_1 = \frac{(1-v)^2 - v^2 A^2 N^2}{4(1-v)^2 A} \tag{0.27}$$

Solving Equation (0.26) for $g[y]$ gives

$$g[y] = \frac{1}{4A}\left(1 - \sqrt{1 - 4n_1 A y}\right)^2 \tag{0.28}$$

Following the steps leading to Equation (0.19), we obtain $n_s$ in the modified model:

$$n_s \simeq N \left( \frac{4(1-v)\left((1-v) + \frac{N v}{\sum_{s'=1}^{\infty}(s')^{1-\delta} n_{s'}}\right)}{\left(\frac{N v}{\sum_{s'=1}^{\infty}(s')^{1-\delta} n_{s'}} + 2(1-v)\right)^2} \right)^s s^{-(5/2-\delta)} \tag{0.29}$$

For $\delta = 0$, $\sum_{s'=1}^{\infty}(s')^{1-\delta} n_{s'} = N$ and hence Equation (0.29) reduces to the result in Equation (0.19) for the original model. For $\delta \neq 0$, it is difficult to solve explicitly for $n_s$. However for small $v$, the dominant dependence on the attack strength $s$ should be $n_s \sim s^{-(5/2-\delta)}$ and hence equivalently $n_s \sim s^{-(2.5-\delta)}$. Although this result is only approximate given that the prefactor now has a highly non-trivial dependence on $s$, numerical simulations show that it is reasonably accurate.

**We can therefore can see that by decreasing $\delta$ from $0.7 \to 0$ (i.e. by increasing the relative fragmentation/coalescence rates of larger attack units) we span the entire spectrum of power-law exponents observed in the Iraq war (see Figure 2) from the initial value of 1.8, up to the current value of approximately 2.5. This effect of decreasing $\delta$ from $0.7 \to 0$ corresponds in our model to a relative increase in the tendency for larger attack units to fragment at each timestep. In other words, decreasing $\delta$ mimics the effect of decreasing the relative robustness or 'lifetime' of larger attack units. A somewhat similar argument applies to Colombia using negative**

**values for $\delta$: by increasing $\delta$ from about $-0.5 \to 0$ we span the spectrum of power-law exponents observed in the Colombia war (see Figure 2) from the initial value of near 3, down to the current value of approximately 2.5. This effect of increasing $\delta$ corresponds in our model to a relative decrease in the tendency for larger attack units to fragment at each timestep. As suggested above, for technical reasons this argument is slightly less satisfactory than for the case of Iraq -- this is because $f[s]$ will now need to have a non-monotonic functional form since it must still normalize to 1 over all $s$.**

Going further, we note that the above theoretical results are consistent with, *and to some extent explain*, the various power-law exponents found for:

(1) Conventional wars. The corresponding power-law exponent 1.8 referred to in the paper for conventional wars, can now be interpreted through our generalized model (with $\delta \approx 0.7$) as a tendency toward building larger, robust attack units with a fixed attack strength as in a conventional army -- as opposed to attack units with rapidly fluctuating attack strengths as a result of frequent fragmentation and coalescence processes. There is also a tendency to form a distribution of attack units with a wider spectrum of attack strengths – this is again consistent with the composition of 'conventional' armies.

(2) Terrorism in G7 countries. The corresponding power-law exponent 1.7 for G7 terrorism can now be interpreted through our generalized model (with $\delta \approx 0.8$) as an even stronger tendency for robust units (e.g. terrorist cells) to form. There is also an increased tendency to form larger units – or equivalently, to operate as part of a large organization.

(3) Terrorism in non-G7 countries. The corresponding power-law exponent 2.5 for non-G7 terrorism can now be interpreted through our model (with $\delta = 0$) as a tendency toward more transient attack units than for G7 terrorism, with attack strengths which are continually evolving dynamically as a result of an on-going fragmentation and coalescence process. Unlike a conventional army, there will be a tendency to form smaller attack units rather than larger ones.

As suggested above, the evolution of the wars in Colombia and Iraq can also be discussed in such terms:

*War in Colombia*. At the beginning of the 1990's, the power-law exponent was very high. Then over the following 15 years, it gradually lowered and has hovered near 2.5 from 1999 onwards. Using our model, the interpretation is that the war at the beginning of the 1990's was such that the guerrillas favored having small attack units. This is possibly because they lacked communications infrastructure, and/or did not feel any safety in larger numbers. The decrease toward the value 2.5, suggests that this has changed – probably because of increased infrastructure and communications, enabling attack units with a wide range of attack strengths to build up.

*War in Iraq*. At the beginning of the war in 2003, the power-law exponent was quite low and was essentially the same value as conventional wars. This is consistent with the war being fought by a conventional Iraqi army against the Coalition forces. There is then a break in this value after a few months (i.e. the conventional war ended) and following this, the power-law exponent gradually rose towards 2.5. This suggests that the insurgents have been increasingly favoring more temporary attack units, with an increasingly rapid fragmentation-coalescence process. This finding could be interpreted as being a result of increased success by the Coalition Forces in terms of forcing the insurgents to fragment. On the other hand, it also means that the Iraq War has now moved to a value of $\alpha$, and hence character, which is consistent with generic non-G7 terrorism and it may therefore be hard to make further progress against them.

Finally we comment in more detail on the generality of the model. As noted by D'Hulst and Rodgers (see preprint cond-mat/9908481 at xxx.lanl.gov) the power-law exponent of 2.5 in this model seems to survive under a range of different generalizations. In particular, if the number of attack units which can coalesce at a given timestep is any number m, as opposed to simply m=2 as in the present version, the power-law exponent remains unchanged at exactly 2.5. D'Hulst and Rodgers suggest that allowing for a variable m, or a different process of fragmentation -- such as allowing some agents to remain connected after a fragmentation process -- will also preserve the value of the power-law exponent around 2.5. We have investigated such robustness further, with the help of Ben Burnett and Alex Dixon at University of Oxford, Department of Physics. Their findings, which are both analytic and numerical, confirm this general result of the robustness of the value 2.5. Indeed, we have even extended the model to account for several co-existing insurgent groups, and also for having one particular insurgent group have a 'home advantage' in clashes. A power-law still emerges, and it still seems to have an exponent $\alpha$ in the range 1.5 to 3.5. Hence one could say that despite the inherent uncertainty concerning the precise microscopic rules which describe a particular conflict, the value of the 2.5 for the power-law exponent will represent a good a priori expected value. Even though the form of the high-s cutoff may depend on the details of the particular conflict, we can conclude that the presence of a power-law with an exponent $\alpha$ of roughly 2.5 over a reasonably wide of range of s values is indicative of something generic in the mechanics of the conflict. In particular, it suggests that the way in which militarily inferior insurgent armies mount attacks against incumbent forces, is generic. This in turn suggests that any group of humans would end up waging war in the same way, if they were to be found fighting in such an asymmetric situation and without any dominant central coordination.

In light of the above discussion and findings, we are confident that the value of 2.5 which is seen to be the current value in Iraq and Colombia, and is also very similar to other modern wars such as Afghanistan, is indeed a meaningful and significant value, and that the underlying common mechanism is indeed one of coalescence and fragmentation by the insurgent army.

**PART 2: Data and methods**

**Data:** For Colombia, we are able to work with the very broad measure of all conflict-related killings plus injuries taken from the CERAC Colombia Conflict Database (CCCD). The CCCD builds on primary source compilations of violent events by Colombian human rights NGO's and from local and national press reports. We distil from this foundation all the clear conflict events, i.e., those that have a military effect and reflect the actions of a group participating in the armed conflict. For each event we record the participating groups, the type of event (massacre, bombing, clash, etc.), the location, the methods used and the number of killings and injuries of people in various categories (guerrillas, civilians, etc.). This data set covers the years 1988-2004 and includes 20,227 events. More information is available at **http://www.cerac.org.co/** . Since the materials found on this website give very detailed descriptions of CCCD, we will not reproduce such details here.

For Iraq we work with the CERAC Integrated Iraq Dataset (CIID). The CIID builds on the event description from three datasets that monitor violence in Iraq: Iraq Body Count (**http://www.iraqbodycount.net/**), iCasualties (**http://iCasualties.org/oif/**) and ITERATE (**http://www.cba.ua.edu/~wenders/**). All three sources contain event data on the Iraq war from its beginning on March 20, 2003. The first two are continually updated whereas ITERATE is updated on an annual basis so at present only extends through the end of 2004. As we discuss below, ITERATE has a very small impact on CIID so the fact that it stops early does not affect the work of this paper.

The Iraq Body Count Project (IBC) monitors the reporting of more than 30 respected online news sources, recording only events covered by at least two of them. For each event IBC logs the date, time, location, target, weapon, estimates of the minimum and maximum number of civilian deaths and the sources of the information.[vii] IBC attaches the most confidence to their figures on the minimum number of killings in each event so the figures in the paper are based on these minimum numbers. However, figure S8 (below) shows that Figure 1 changes very little if we substitute the maximum number of killings for the minimum number of killings. The concept of civilian is broad, including, for example, policemen. The list of events, posted online, covers the full range of war activity, including suicide bombings, roadside bombings, US air strikes, car bombs, artillery strikes and individual assassinations.

The IBC data has two principle drawbacks which need to be addressed in order for the reader to have confidence in our results. First, some lines in the IBC spreadsheet contain entries that are not proper events. The most important entries of this form are based on reports from morgues around Iraq. For example, entry x355a lists 26 deaths between May 1, 2003 and May 31, 2003 described as "Violent deaths recorded at the provincial morgue of Karbala". The following hypothetical calculation illustrates how IBC handles these entries. The Karbala morgue actually reports a higher figure, say 39 violent deaths for May of 2003. However, IBC already has two events in Karbala for May of 2003, a car bomb killing 4 and a suicide bombing killing 5. It is likely that these 9 deaths are included among the 39 violent deaths recorded by the Karbala morgue so IBC subtracts them off, leaving 30. In addition, the murder rate in Karbala before the war was 4 per months so we might expect that 4 out of the remaining violent deaths would have happened even without the war. IBC subtracts off these 4 leaving the figure of 26 which is the one they enter into the database. This procedure is reasonable on its own terms, however deeply problematic for our purposes in this paper for two reasons. First, most of the deaths in entries of this

---

[vii] IBC also records injuries but does not post this information on its open website.

form are likely to have occurred as single homicides since larger conflict events would be likely to have their own entry. Clearly we would not wish to treat 26 individual homicides as one event in which 26 people were killed. Second, most of the killings in events of this form are more tied to crime than to the conflict directly and we prefer to focus on a narrow definition of conflict killing. For these reasons we delete from the IBC database entries of this nature. We provide a list of events we deleted from IBC in Part 4 of the *Appendices*.

The second drawback of IBC is that it measures only civilian deaths, albeit with a wide concept of civilian. Therefore, to get of fuller picture of the conflict we have added in events from iCasualties in which coalition military personnel and contractors are killed in conflict events (but not in accidents). This is a highly reliable source as the military services keep solid records on the fate of their own personnel. Finally, as a check for coverage of IBC and iCausalties we also integrated events from ITERATE, which is a global terrorism database that records terrorism events of international significance.

This integration required careful matching of events between the three sources to avoid double counting. The following Venn diagrams give the results of this matching work, with event counts and numbers of killings accounted for by these events in parentheses below. They show that most of the deaths in CIID come from IBC alone but that iCasualties does make a significant contribution. The overlap between IBC and iCasualties is small, since they are measuring different things. However, there is some overlap because sometimes both military personnel and civilians are killed in a single incident. The impact of adding ITERATE into CIID is negligible, indicating that IBC and iCasualties give very full coverage of the Iraq war.

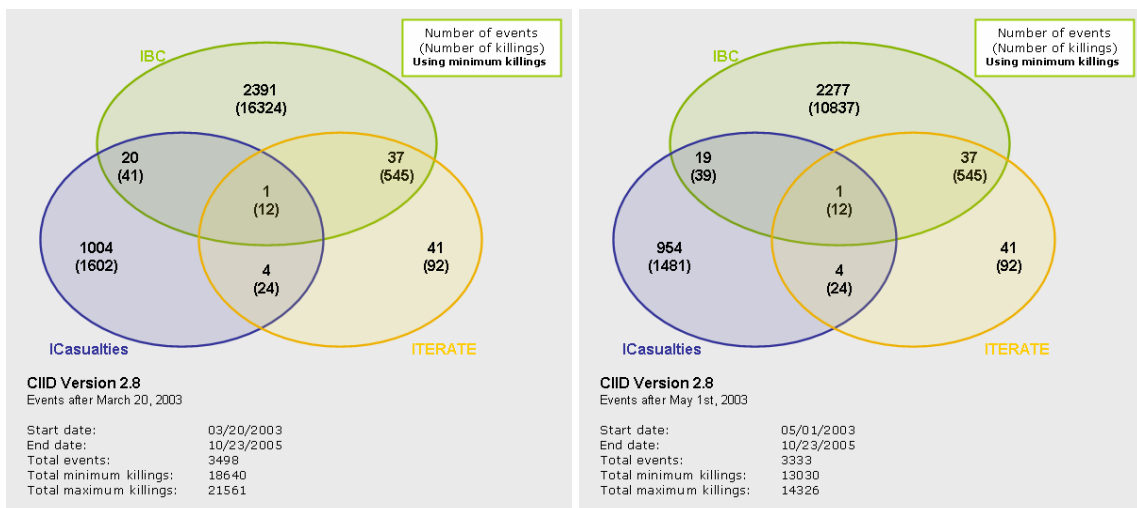

**Figure S2: Venn diagrams showing the sources behind the CIID database for 03/20/03 – 10/23/05 and 5/1/03 – 10/23/05. Overlaps indicate that the same event is reported in multiple sources. The diagram gives the number of events with various degrees of overlap and (in parentheses) the minimum estimate of the number of killings accounted for by these events.**

The Iraq war breaks naturally into two phases. In the first phase, roughly from March 20, 2003 to April 30, 2003 the conflict had the character of a conventional war, marked by heavy aerial bombardments and a big push across Iraq by coalition land forces. From May 1, 2003 the war took on an "irregular" or "insurgent" character. Accordingly, in figure 1 we used just the 3,333 events from May 1, 2003 until October 23, 2005 to ensure maximum comparability to the Colombian data. However, figure S3 (below) shows that we get similar results if we use the combined data from both phases. Figure 2, showing the evolution of both conflicts, uses all the Iraq data beginning from March 20, 2003 to October 23, 2005.

We note that the largest level of killings in events shown in Figure 1, according to the CIID database, occurred in Baghdad in 8/31/05 (965 killed) and in Fallujah where there were two multi-day sieges: the first between 11/8/04 - 11/30/04 (581 killed) and the second between 4/5/04 and 4/30/04 (572 killed) which we have necessarily had to treat as single events. It turns out that even if we choose to exclude these three events, there is only a small impact on the parameter estimation in Figure 1, raising $x_{\min}$ from 7 to 8 and $\alpha$ from 2.31 to 2.42. Hence we are confident that our results are not distorted by the method chosen to characterize these particular large events. We also note that the Colombia curve is less noisy at high $x$ because that war has more events in this region.

The CERAC Integrated Afghanistan Database (CIAD) is constructed very similarly to CIID. Two of the basic sources, iCasualties and ITERATE, are the same as in CIID, and provide 126 and 29 events respectively to CIAD. The only difference is that IBC is replaced by data on Afghanistan provided by Marc Herold (**http://pubpages.unh.edu/~mwherold/**). This project provided the original model for IBC and so has a very similar nature. Herold monitors a large number of English language sources on the conflict in Afghanistan and provides a list of events with minimum and maximum killings on his web site. (Figure S9 shows that our results are robust to whether we use minimum killings or maximum killings.) The data comes in two separate files, the Afghanistan Daily Count database (ADC) with 763 events and the Day by Day Chronicle (DBDC) with 608 events. ADC covers killings of Afghan civilians attributed to coalition forces during the period October 7, 2001 to June 3, 2003. DBDC covers May 31, 2003 to July 31, 2004 and covers both civilian and military casualties, although this expansion has only a small impact on CIAD which already integrates information from iCasualties. CIAD holds a total of 1430 registries between October 7, 2001 and July 31, 2004.

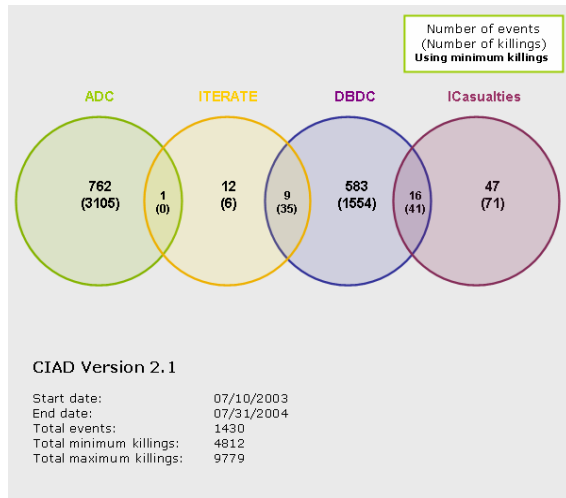

**Figure S3: Venn diagrams showing the sources used to build the CIAD database for 10/07/03 – 07/31/04. Overlaps indicate that the same event is reported in two sources. The diagram gives the number of events and (in parentheses) the minimum estimate of the number of killings accounted for by these events.**

**Methods:** Let $X$ be a random variable that follows a discrete power law for values greater or equal to $x_{min}$. The probability that $X$ takes a specific value $k$ conditional on $X \geq x_{min}$, denoted by $p(\cdot)$, is:

$$p(X = k) = \frac{k^{-\alpha}}{\zeta(\alpha, x_{min})} \text{ for all } k \geq x_{min} \qquad (0.30)$$

where $\zeta(\alpha, x_{min}) = \sum_{i=x_{min}}^{\infty} i^{-\alpha}$ is the incomplete Riemann zeta function.[5] Suppose we take $n$ i.i.d. draws from this distribution which we call $x_1, x_2, ..., x_n$. Conditional on a fixed value of $x_{min}$, which we call $x_{min}^0$, the maximum likelihood estimator of the $\alpha$ parameter is the $\alpha$ value that maximizes the log-likelihood function:

$$\ell(\alpha \mid x_1,\ldots,x_n; x_{\min}^0) = -\sum_{i=1}^{n}\left(\alpha \ln x_i + \ln\left[\zeta(\alpha, x_{\min}^0)\right]\right) \qquad (0.31)$$

To select the $x_{\min}$ parameter we use a grid search, which is based on the following procedure:

1. Create a list of $K$ feasible values for $x_{\min}$, denoted $x_{\min}^1 = 1, x_{\min}^2 = 2,\ldots, x_{\min}^K = K$. For each $x_{\min}^k$, $k = 1,\ldots, K$, associate all the values of $X$ that are greater than or equal to $x_{\min}^k$;[viii]

2. Compute the maximum likelihood estimator for the $\alpha$ parameter according to equation (0.31) for each combination of $x_{\min}^k$ and $X$'s above $x_{\min}^k$, with $k = 1,\ldots, K$. This process yields a total of $K$ estimates of $\alpha$;

3. For each combination of $x_{\min}^k$ and $X$'s above $x_{\min}^k$, with $k = 1,\ldots, K$, calculate the Kolmogorov-Smirnov (KS) goodness-of-fit statistic, which gives the maximum distance between the estimated theoretical distribution and the empirical distribution. Note that for each subset the empirical and theoretical distributions start at $X = x_{\min}^k$;

4. Select $x_{\min}$ as the value of $x_{\min}^k$ that minimizes the KS test statistic and the $\alpha$ estimate as the $\alpha$ value associated with this $x_{\min}$.

5. Compute confidence bands for the $\alpha$ parameter estimate through bootstrapping, which proceeds as follows. Generate a new dataset where each point is a random draw from the empirical distribution of the (observed) data given $x_{\min}$. This draw contains a number of data points exactly equal to the number of points in the original estimated power law. Repeat this process to create 2500 such datasets. Keeping $x_{\min}$ fixed we then estimate the

---

[viii] For all three cases in this paper we used $K = 50$.

$\alpha$ parameter for each dataset by maximum likelihood estimation as defined in equation (0.31) and thus we obtain the distribution of the 2500 "bootstrapped" alphas. The limits of the 95% confidence interval are given by the percentiles 2.5 and 97.5 of this distribution, respectively.

An alternative method for calculating confidence intervals on our $\alpha$ estimates uses the well-known property of asymptotic normality of the maximum likelihood estimators. We prefer the bootstrap approach because it can display the stochastic properties of the maximum likelihood estimator well, even in small samples when there can be significant departures from normality. We find no difference in the two approaches for our large Colombia dataset but the two approaches do differ for Iraq and Afghanistan.

Next we test the null hypothesis that the data follows the power-law estimated according to the above procedure. First, we randomly draw 2000 samples, each generated from the theoretical power-law distribution given by the estimated $\alpha$ and $x_{min}$ and each with the same number of data points as in the observed data for $x \geq x_{min}$. We compute the KS statistic, comparing these samples to the theoretical distribution to obtain the distribution of the KS test statistic, conditional on the true data generation process being a power law with the estimated $\alpha$ and $x_{min}$. From this we calculate the probability that a power law with this estimated $\alpha$ and $x_{min}$ would generate a KS statistic bigger than the one we found in Step 4 of the above procedure on the real data. For all three conflicts, these *p* values always come out far above any standard critical level.

To further test the reliability of our result, we use the same Monte Carlo procedure just described above to compare our data against the lognormal distribution. This is a natural comparator, as the lognormal shares some features of the power law distribution: they both exhibit fat tails; they are both defined only for positive values; and most importantly, the lognormal distribution could possibly resemble a straight line for some ranges of $X^1$. Again, we use the KS goodness-of-fit statistic to perform a test of rejection of the null of lognormality; in this case the test is calculated over all casualty levels, above and below $x_{min}$, to see whether the whole sample follows a lognormal distribution. Results of these tests with the associated *p*-values are presented in Table T2 and suggest that we cannot reject with any reasonable degree of confidence the hypothesis that our data does not follow a lognormal distribution. In other words we can be confident that the data follows a power-law distribution, and does not follow a lognormal one.

For Figure 2, we calculated the $\alpha$'s for a sequence of sliding time windows for both Iraq and Colombia. This work follows the time evolution of the power-law coefficient $\alpha$ through monthly estimation, using fixed-width time windows within each conflict to determine the sample-defined path of the estimated $\alpha$. For Colombia each window was 800 days long, and we slid it forward one month at a time. For Iraq we used 400 day time-windows which we also slid forward one month at a time. Events in the Colombia conflict are roughly half as frequent as in Iraq, so we took advantage of the long run of Colombia data and built longer time-windows for Colombia than we did for Iraq. Figure 2 also displays symmetric confidence bands at 95%.[ix] As a further check, we then repeated the calculation of $\alpha$ for several different sizes of the time-windows and for the overlap interval which slides it forward in time. None of these modifications affected our main results or conclusions. Figure S6 show results with a 2,500 day time-window sliding every 60 months for Colombia, and a 365 day time-window displaced every 8 months for Iraq. Figure S7

---

[ix] For these confidence intervals, we used the asymptotic normality of the ML estimator rather than bootstrapping. The latter approach makes extreme computational demands, given the number of confidence intervals to compute.

shows a moving time window of two years displaced every year for Colombia, and a 250 day interval displaced every 150 days for Iraq. As can be seen, our results are essentially unchanged by these variations. Furthermore, our results (whose details are available from the authors upon request) are statistically robust to the exclusion of high leverage points and extreme value observations.

It is instructive to compare our discrete maximum likelihood (DML) estimator to Ordinary Least Squares (OLS) and continuous maximum likelihood (CML) estimators (see Ref. [1] for a discussion regarding the different types of estimators). Table T1 shows the results of numerical Monte Carlo simulations using the Mean Squared Error (MSE) for different sample sizes and varying $x_{min}$ for a theoretical power-law distribution with a fixed $\alpha = 2.5$.[x] An estimator is better than other if it has a lower MSE. The DML estimator is the best one of the ones here compared.[xi]

| | | Mean Squared Error (MSE) | | | | |
|---|---|---|---|---|---|---|
| Estimator | Obs | $x_{min}=1$ | $x_{min}=2$ | $x_{min}=5$ | $x_{min}=10$ | $x_{min}=50$ |
| DML | 100 | 0.0320 | 0.0255 | 0.0240 | 0.0238 | 0.0236 |
| DML | 500 | 0.0057 | 0.0047 | 0.0045 | 0.0045 | 0.0044 |
| DML | 1000 | 0.0028 | 0.0024 | 0.0023 | 0.0022 | 0.0022 |
| DML | 5000 | 0.0006 | 0.0005 | 0.0005 | 0.0005 | 0.0004 |
| | | | | | | |
| CML | 100 | 5.1161 | 0.7871 | 0.1230 | 0.0513 | 0.0263 |
| CML | 500 | 4.0614 | 0.6242 | 0.0764 | 0.0214 | 0.0054 |
| CML | 1000 | 3.9700 | 0.6083 | 0.0715 | 0.0181 | 0.0030 |

---

[x] Simulations were performed with random samples of 100, 500, 1,000 and 5,000 and with 4,000 replications.

[xi] The MSE represents the expected squared distance of the estimator from the population value; i.e., as the MSE is the sum of the squared bias and the variance of the estimator it is a weighted summary of the bias and its efficiency, see Wooldridge, Jeffrey M., 2003, Introductory Econometrics: A Modern Approach, 2 ed., South-Western Publisher of Thomson Learning.

| Estimator | Obs | $x_{min}=1$ | $x_{min}=2$ | $x_{min}=5$ | $x_{min}=10$ | $x_{min}=50$ |
|---|---|---|---|---|---|---|
| CML | 5000 | 3.8808 | 0.5927 | 0.0670 | 0.0152 | 0.0011 |
|  |  |  |  |  |  |  |
| OLS | 100 | 0.4125 | 0.3595 | 0.3349 | 0.3271 | 0.2915 |
| OLS | 500 | 0.3206 | 0.2920 | 0.2770 | 0.2635 | 0.1310 |
| OLS | 1000 | 0.2986 | 0.2775 | 0.2609 | 0.2345 | 0.0681 |
| OLS | 5000 | 0.2576 | 0.2416 | 0.1944 | 0.1213 | 0.0183 |
| BIAS |  |  |  |  |  |  |
| Estimator | Obs | $x_{min}=1$ | $x_{min}=2$ | $x_{min}=5$ | $x_{min}=10$ | $x_{min}=50$ |
| DML | 100 | 0.0216 | 0.0160 | 0.0145 | 0.0144 | 0.0156 |
| DML | 500 | 0.0039 | 0.0030 | 0.0027 | 0.0028 | 0.0040 |
| DML | 1000 | 0.0027 | 0.0022 | 0.0020 | 0.0020 | 0.0033 |
| DML | 5000 | 0.0007 | 0.0006 | 0.0005 | 0.0006 | 0.0020 |
|  |  |  |  |  |  |  |
| CML | 100 | 2.1124 | 0.8147 | 0.2797 | 0.1384 | 0.0390 |
| CML | 500 | 1.9897 | 0.7761 | 0.2608 | 0.1237 | 0.0270 |
| CML | 1000 | 1.9797 | 0.7728 | 0.2594 | 0.1227 | 0.0263 |
| CML | 5000 | 1.9674 | 0.7684 | 0.2572 | 0.1209 | 0.0248 |
|  |  |  |  |  |  |  |
| OLS | 100 | -0.5182 | -0.4524 | -0.4189 | -0.4094 | -0.3885 |
| OLS | 500 | -0.4556 | -0.4084 | -0.3846 | -0.3734 | -0.2722 |
| OLS | 1000 | -0.4299 | -0.3916 | -0.3702 | -0.3530 | -0.1838 |
| OLS | 5000 | -0.3864 | -0.3623 | -0.3283 | -0.2649 | -0.0304 |

**Table T1** Mean Square Error and Bias of the three feasible estimators for the estimation of the $\alpha$ power-law parameter for varying levels of $x_{min}$ values and sample sizes in Monte Carlo simulations with 4,000 replications.

Furthermore, Table T1 also shows the bias of each one of the possible estimators to use. We find that the DML estimator has the lowest (upward) bias, that this bias falls rapidly with larger sample sizes and does not depend on the $x_{min}$ value. On the other hand, the CML estimator has the largest (upward) bias of all, although for high $x_{min}$ values this bias becomes rather small. The bias of the CML estimator does not fall as the sample size grows. Finally, we find that the OLS estimator has a (downward) bias which does not depends on the sample size or the $x_{min}$ values. Figure S4 shows the distributions of the three estimators for a Monte Carlo simulation with a sample size of 500 and an $x_{min}=5$.

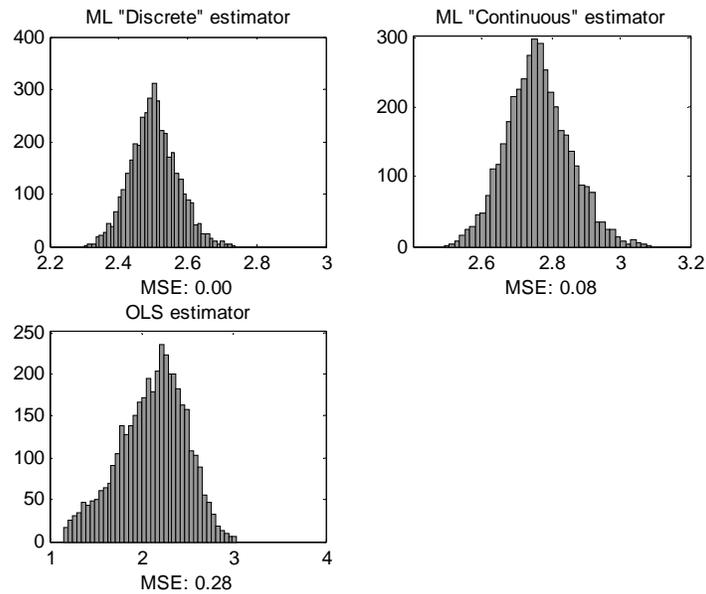

**Figure S4: Simulated Monte Carlo distributions of the estimators of the $\alpha$ power-law parameter for the Discrete Maximum Likelihood (DML), Continuous Maximum Likelihood (CML) and Ordinary Least Squares (OLS) estimators, with a sample size of 500 and an $x_{\min} = 5$. The DML estimator shows the lowest Mean Squared Error (MSE) and Bias.**

PART 3: Tables and figures confirming the reliability of our results

| Estimates of power-law coefficients for the entire time-series | | | | | | | |
|---|---|---|---|---|---|---|---|
| | Country | $\alpha$ | $\alpha$ Lower Confidence Band | $\alpha$ Upper Confidence Band | $x_{\min}$ | Significance of KS test for Power Law dist. | Significance of KS test for Lognormal dist. |
| K | Colombia | 2.9622 | 2.8847 | 3.0465 | 5 | 0.4860 | <0.001 |
| I | Colombia | 2.7557 | 2.6345 | 2.8863 | 6 | 0.7780 | <0.001 |
| KI | Colombia | 2.7896 | 2.7138 | 2.8722 | 7 | 0.5550 | <0.001 |
| $K_{min}$ | Iraq | 2.3135 | 2.1765 | 2.4766 | 7 | 0.9970 | <0.001 |
| $K_{max}$ | Iraq | 2.1612 | 2.1025 | 2.2313 | 3 | 0.4280 | <0.001 |
| $K_{min}$ | Afghanistan | 2.4462 | 2.2117 | 2.8053 | 13 | 0.9050 | <0.001 |
| $K_{max}$ | Afghanistan | 2.1629 | 2.0266 | 2.3387 | 10 | 0.5410 | 0.0448 |

**Table T2** Here we use a variety of measures of the impact of violent events to show that our results vary little depending on the choice of violence variables. For Colombia we try killings (K), injuries (I) and, as in the paper, killings plus K and I. For Iraq and Afghanistan we use minimum killings, $K_{min}$, as in the paper, and also maximum killings, $K_{max}$. Our $\alpha$ parameter estimate is the maximum likelihood estimator for a discrete power law. For further discussion, see PART 2 of these *Appendices*.

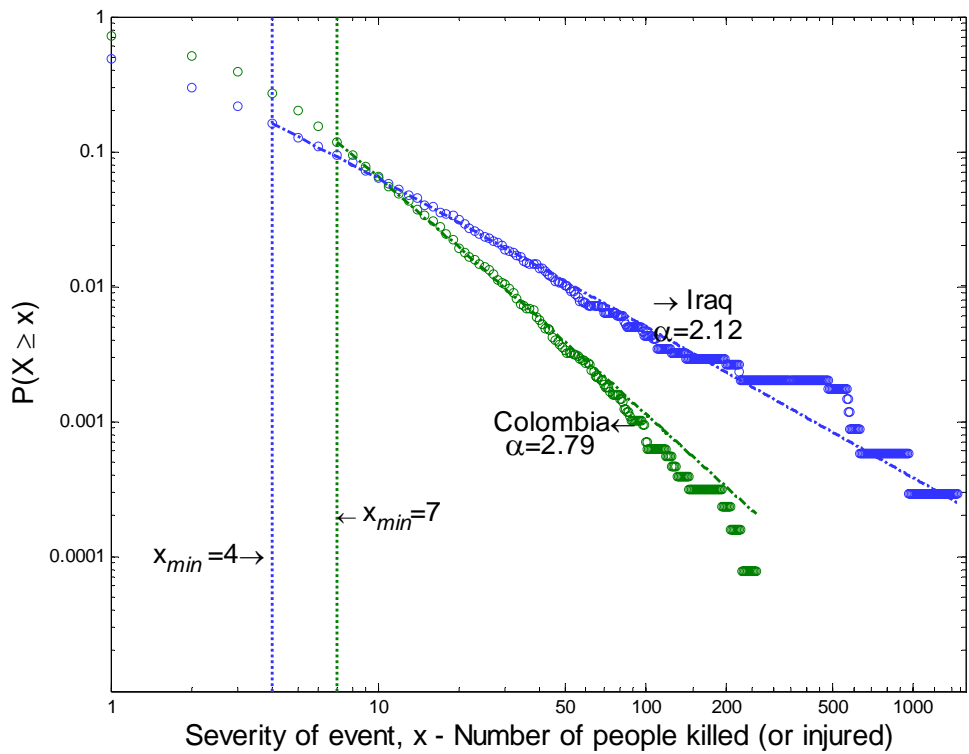

**Figure S5** This is the same as Figure 1 but with the whole Iraq war beginning 20/03/03 rather than 01/05/03. The results are very similar but Iraq now has a lower $x_{\min}$ (4) and a slightly lower alpha (2.12 vs. 2.31).

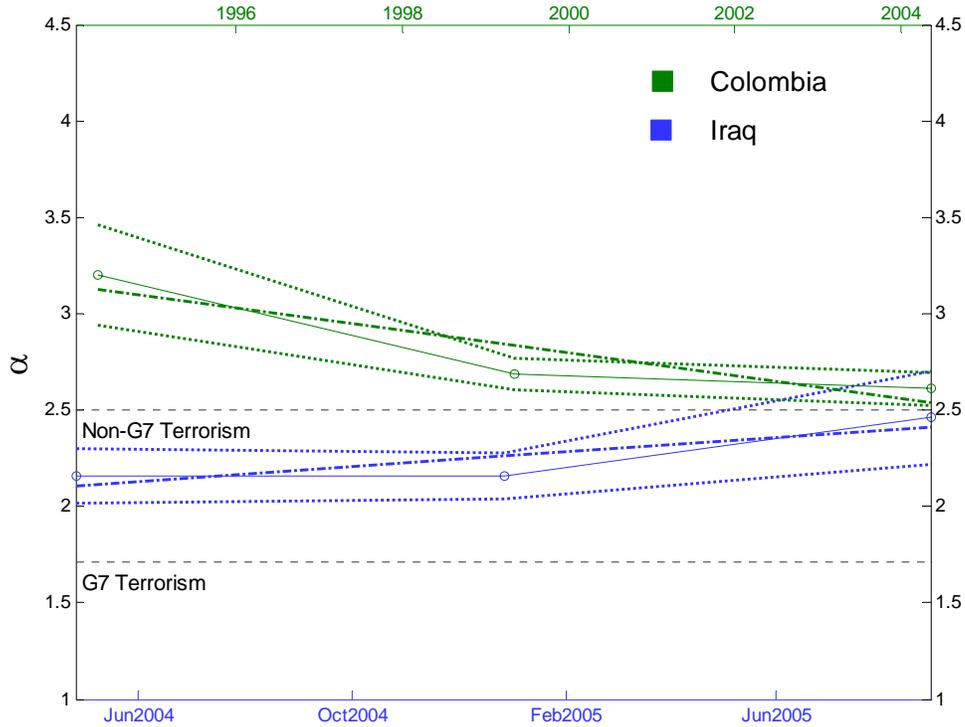

**Figure S6** This gives the variation through time of the power-law coefficient $\alpha$ for each war, using much longer time-windows than those used in Figure 2 of the paper. For Colombia we use three 2,500 day intervals displaced by 60 months. For Iraq there are three 365 day intervals displaced every 8 months. Despite this change in size of the windows and how they slide across time, both curves do seem to be heading toward 2.5 as in Figure 2 of the paper. For further discussion, see PART 2 of these *Appendices*.

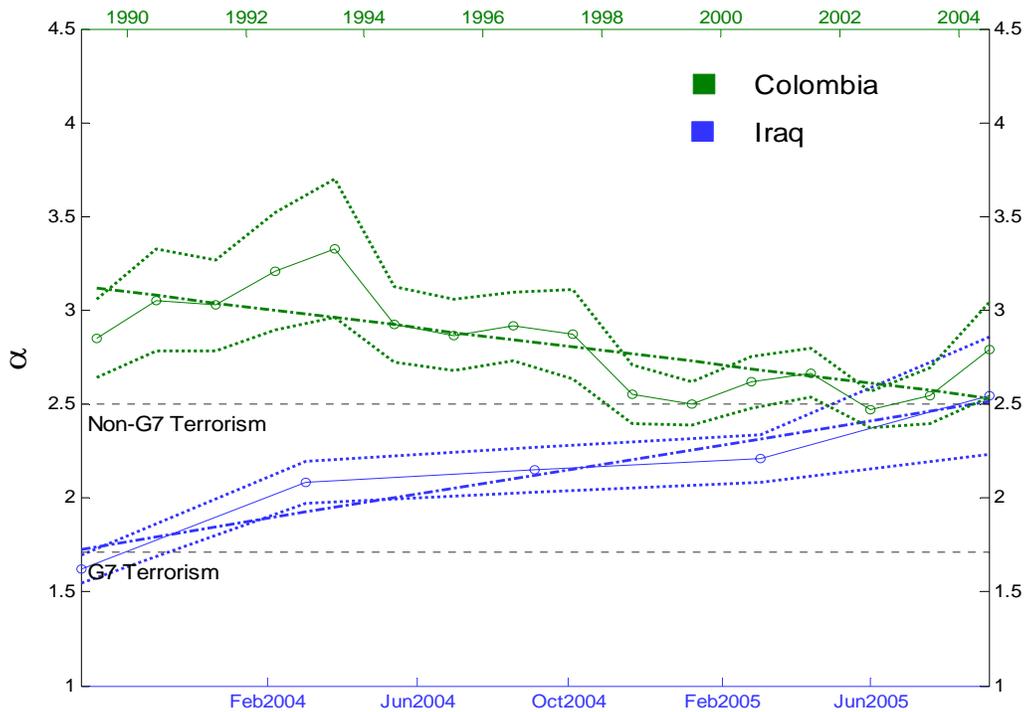

**Figure S7** This gives the variation through time of the power-law coefficient $\alpha$ for two-year intervals displaced every year for Colombia and 250 day intervals displaced every 6 months for Iraq. Again, they both seem to be moving toward 2.5, as in Figure 2 of the paper. For further discussion, see PART 2 of these *Appendices*.

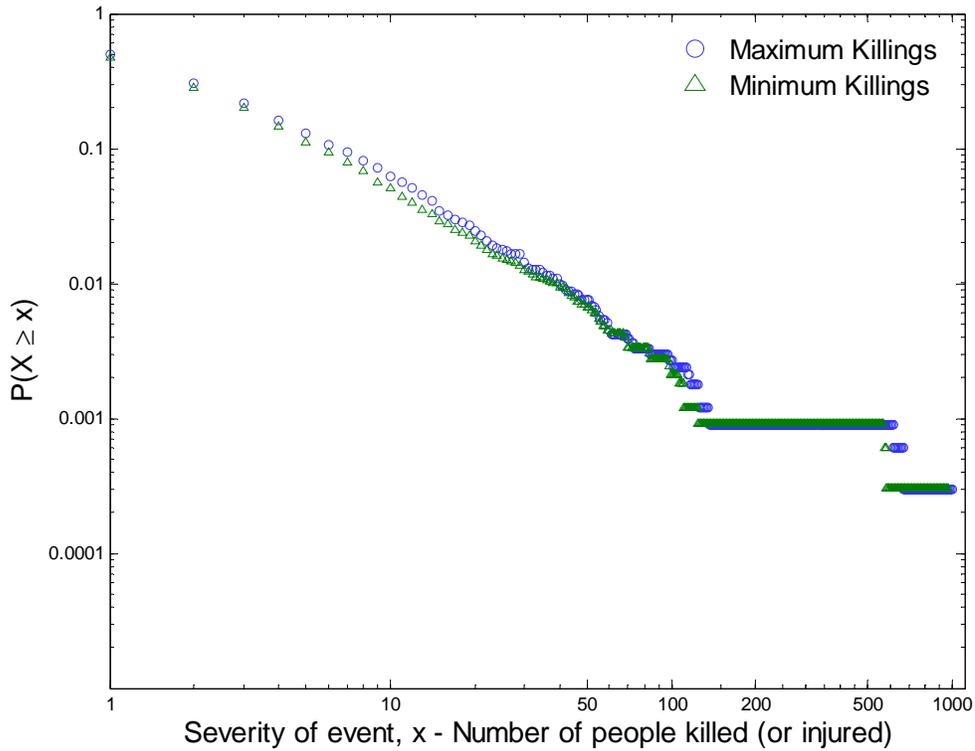

**Figure S8** Log-Log plots of cumulative distributions $P(X \geq x)$ describing events greater than $x$ for the minimum possible value and maximum possible value of each event in the Iraq dataset. The results are very much the same across the two measures. For further discussion, see PART 2 of these *Appendices*.

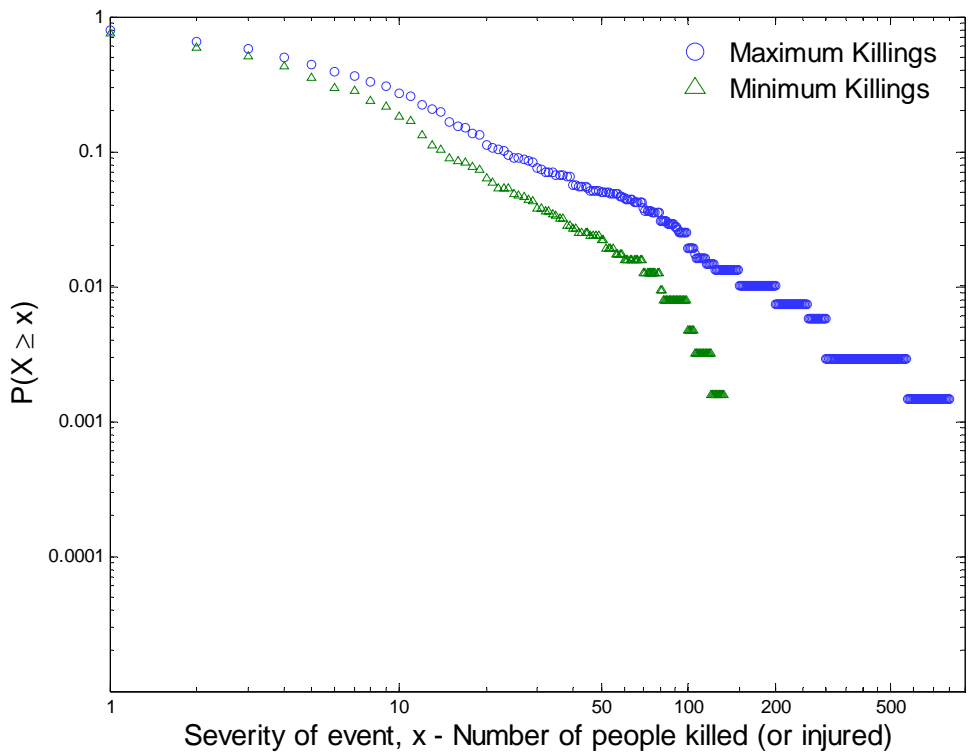

**Figure S9** Log-Log plots of cumulative distributions $P(X \geq x)$ describing events greater than $x$ for the minimum possible value and maximum possible value of each event in the Afghanistan dataset. The main features of the results, such as the slopes, are similar for the two measures. Most importantly, as noted in Table T2, we find that (a) we cannot reject a power-law in either case, and (b) we can reject lognormality in both cases. For further discussions, see PART 2 of these *Appendices*.

Finally we comment on an important advantage of focusing on the value of $\alpha$ in order to characterize wars, as we have done in this paper. This advantage concerns possible over- or under-reporting of war casualties. In particular, it can be shown that $\alpha$ is insensitive to systematic over-reporting or under-reporting of casualties. This is because any systematic multiplication of the raw numbers by some constant factor has no effect on the $\alpha$ value (i.e. slope) which emerges from the log-log plot for a power-law. Such systematic multiplication just rescales the intercept, leaving the slope unchanged.

PART 4: List of IBC events deleted in forming CIID

**Table T3** Detail of events excluded from IBC by reason of exclusion

Reason 1:
Morgue reports aggregate many killings over extended period: many may be single homicides, possibly criminal rather than conflict

| Start Date | City | Killings Min. | Killings Max |
|---|---|---:|---:|
| 4/14/03 | Baghdad | 0 | 41 |
| 5/1/03 | Unknown | 26 | 26 |
| 5/1/03 | Unknown | 49 | 49 |
| 5/1/03 | Unknown | 15 | 16 |
| 5/1/03 | Unknown | 275 | 291 |
| 6/1/03 | Karbala | 27 | 27 |
| 6/1/03 | Kirkuk | 34 | 35 |
| 6/1/03 | Tikrit | 23 | 22 |
| 6/1/03 | Baghdad | 334 | 349 |
| 7/1/03 | Karbala | 15 | 16 |
| 7/1/03 | Kirkuk | 50 | 51 |
| 7/1/03 | Tikrit | 5 | 5 |
| 7/1/03 | Baghdad | 438 | 467 |
| 8/1/03 | Karbala | 25 | 25 |
| 8/1/03 | Kirkuk | 35 | 36 |
| 8/1/03 | Tikrit | 2 | 1 |
| 8/1/03 | Baghdad | 470 | 492 |
| 9/1/03 | Baghdad | 362 | 367 |
| 9/1/03 | Karbala | 12 | 14 |
| 9/1/03 | Kirkuk | 26 | 28 |
| 9/1/03 | Tikrit | 12 | 13 |
| 10/1/03 | Karbala | 10 | 2 |
| 10/1/03 | Kirkuk | 21 | 21 |
| 10/1/03 | Tikrit | 8 | 10 |
| 10/1/03 | Baghdad | 267 | 272 |
| 11/1/03 | Unknown | 26 | 33 |
| 11/1/03 | Baghdad | 256 | 267 |
| 12/1/03 | Karbala | 28 | 28 |
| 12/1/03 | Kirkuk | 20 | 21 |
| 12/1/03 | Tikrit | 13 | 13 |
| 12/1/03 | Baghdad | 287 | 296 |
| 1/1/04 | Karbala | 94 | 94 |
| 1/1/04 | Kirkuk | 12 | 12 |
| 1/1/04 | Tikrit | 6 | 8 |
| 1/1/04 | Baghdad | 268 | 277 |
| 2/1/04 | Karbala | 18 | 18 |
| 2/1/04 | Baghdad | 210 | 220 |
| 3/1/04 | Karbala | 217 | 221 |

Reason 1:
Morgue reports aggregate many killings over extended period: many may be single homicides, possibly criminal rather than conflict

| Start Date | City | Killings Min. | Killings Max |
|---|---|---|---|
| 3/1/04 | Kirkuk | 14 | 14 |
| 3/1/04 | Tikrit | 9 | 11 |
| 3/1/04 | Baghdad | 317 | 334 |
| 4/1/04 | Baghdad | 232 | 236 |
| 4/1/04 | Karbala | 50 | 52 |
| 4/1/04 | Kirkuk | 13 | 13 |
| 4/1/04 | Tikrit | 9 | 10 |
| 5/1/04 | Baghdad | 295 | 341 |
| 6/1/04 | Baghdad | 375 | 418 |
| 7/1/04 | Baghdad | 364 | 413 |
| 8/1/04 | Unknown | 283 | 323 |
| 9/1/04 | Unknown | 310 | 353 |
| 10/1/04 | Baghdad | 394 | 435 |
| 11/1/04 | Unknown | 308 | 349 |
| 12/1/04 | Baghdad | 376 | 432 |
| 1/1/05 | Baghdad | 355 | 402 |
| 2/1/05 | Baghdad | 351 | 389 |
| 3/1/05 | Baghdad | 185 | 204 |
| Subtotal | | 8236 | 8913 |

Reason 2:

Police reports aggregate many killings over extended period: many may be single homicides, possibly criminal rather than conflict

| Start Date | City | Killings Min. | Killings Max |
|---:|---|---:|---:|
| 5/1/03 | Baghdad | 16 | 16 |
| 6/1/03 | Baghdad | 18 | 19 |
| 7/1/03 | Baghdad | 20 | 21 |
| 8/1/03 | Baghdad | 25 | 26 |
| 9/1/03 | Baghdad | 17 | 18 |
| 10/1/03 | Baghdad | 15 | 16 |
| 11/1/03 | Baghdad | 14 | 15 |
| 12/1/03 | Baghdad | 10 | 10 |
| Subtotal | | 135 | 141 |

Reason 3:

Accidents

| Start Date | City | Killings Min. | Killings Max |
|---:|---|---:|---:|
| 3/23/04 | Balad | 1 | 1 |
| Subtotal | | 1 | 1 |

Reason 4:

Events before May 1st, 2003

| Start Date | City | Killings Min. | Killings Max |
|---:|---|---:|---:|
| 1/1/03 | Qurnah | 1 | 1 |
| 1/6/03 | Amarah | 2 | 2 |
| 2/10/03 | Basra | 2 | 2 |
| 3/2/03 | Basra | 6 | 6 |
| 3/5/03 | Anbar | 3 | 3 |
| 3/15/03 | Unknown | 1 | 1 |
| 3/20/03 | Zubair | 4 | 60 |
| 3/20/03 | Baghdad | 0 | 24 |
| 3/20/03 | Baghdad | 567 | 978 |
| 3/20/03 | Nassiriya | 633 | 633 |
| 3/20/03 | Nassiriya | 226 | 240 |
| 3/20/03 | Unknown | 30 | 30 |
| 3/20/03 | Baghdad | 200 | 200 |

Reason 4:

Events before May 1st, 2003

| Start Date | City | Killings Min. | Killings Max |
|---|---|---:|---:|
| 3/20/03 | Baghdad | 1473 | 2000 |
| 3/20/03 | Najaf | 224 | 358 |
| 3/20/03 | Basra | 142 | 200 |
| 3/20/03 | Najaf, Karba, Mosul, Samawa, Madain, Diwaniyah, Kut, Tikrit | 484 | 445 |
| 3/20/03 | Baghdad | 22 | 22 |
| 3/20/03 | Rutba | 1 | 1 |
| 3/21/03 | Baghdad | 0 | 3 |
| 3/21/03 | Umm Qasr | 2 | 2 |
| 3/22/03 | ImAnas | 1 | 1 |
| 3/22/03 | Mosul | 4 | 4 |
| 3/22/03 | Nassiriya | 12 | 12 |
| 3/22/03 | Basra | 50 | 77 |
| 3/22/03 | Tikrit | 4 | 5 |
| 3/22/03 | Kurdistan | 57 | 100 |
| 3/23/03 | Najaf | 3 | 8 |
| 3/23/03 | Rutbah | 5 | 5 |
| 3/23/03 | Babel | 30 | 30 |
| 3/23/03 | Basra | 14 | 14 |
| 3/23/03 | Karba | 10 | 10 |
| 3/23/03 | Nassiriya | 10 | 10 |
| 3/24/03 | Baghdad | 5 | 5 |
| 3/24/03 | Baghdad | 5 | 5 |
| 3/25/03 | Ash Shatra | 2 | 2 |
| 3/25/03 | Nassiriya | 2 | 2 |
| 3/26/03 | Rutbah | 2 | 2 |
| 3/26/03 | Baghdad | 14 | 14 |
| 3/26/03 | Baghdad | 21 | 21 |
| 3/27/03 | Missan | 2 | 2 |
| 3/27/03 | Mosul | 2 | 50 |
| 3/27/03 | Waset | 2 | 2 |
| 3/27/03 | Baghdad | 7 | 7 |
| 3/27/03 | Babel | 26 | 26 |
| 3/27/03 | Karba | 11 | 11 |
| 3/27/03 | Hillah | 78 | 201 |
| 3/27/03 | Najaf | 26 | 26 |
| 3/28/03 | Baghdad | 34 | 62 |
| 3/28/03 | Anbar | 28 | 28 |
| 3/28/03 | Babel | 3 | 3 |
| 3/28/03 | Baghdad | 6 | 6 |
| 3/28/03 | Karba | 6 | 6 |
| 3/28/03 | Najaf | 35 | 35 |
| 3/29/03 | Unknown | 1 | 1 |

Reason 4:

Events before May 1st, 2003

| Start Date | City | Killings Min. | Killings Max |
|---|---|---|---|
| 3/29/03 | Janabiin | 20 | 20 |
| 3/30/03 | Baghdad | 15 | 15 |
| 3/31/03 | Baghdad | 6 | 6 |
| 3/31/03 | Mosul | 21 | 21 |
| 3/31/03 | Hillah | 15 | 15 |
| 3/31/03 | Hillah | 24 | 24 |
| 3/31/03 | Najaf and Karba | 11 | 11 |
| 3/31/03 | Baghdad | 24 | 24 |
| 4/1/03 | Baghdad | 1 | 1 |
| 4/1/03 | Shatra | 1 | 1 |
| 4/1/03 | Hillah | 33 | 33 |
| 4/2/03 | Baghdad | 43 | 43 |
| 4/2/03 | Baghdad | 5 | 5 |
| 4/3/03 | Baghdad | 10 | 16 |
| 4/3/03 | Baghdad | 27 | 27 |
| 4/3/03 | Basra | 42 | 51 |
| 4/3/03 | Karba | 5 | 5 |
| 4/3/03 | Najaf | 0 | 40 |
| 4/4/03 |  | 17 | 17 |
| 4/4/03 | najaf | 7 | 7 |
| 4/4/03 | Baghdad | 6 | 6 |
| 4/5/03 | Karba | 1 | 1 |
| 4/5/03 | Baghdad | 22 | 22 |
| 4/5/03 | Rashidiya | 85 | 85 |
| 4/5/03 | Basra | 17 | 17 |
| 4/6/03 | irbil | 1 | 1 |
| 4/6/03 | Baghdad | 15 | 15 |
| 4/6/03 | karbala | 35 | 35 |
| 4/7/03 | Baghdad | 2 | 2 |
| 4/7/03 | Baghdad | 9 | 14 |
| 4/7/03 | Baghdad | 11 | 11 |
| 4/7/03 | Baghdad | 4 | 4 |
| 4/7/03 | Baghdad | 3 | 3 |
| 4/8/03 | Baghdad | 1 | 1 |
| 4/8/03 | Baghdad | 2 | 2 |
| 4/8/03 | Baghdad | 35 | 35 |
| 4/8/03 | Baghdad | 13 | 13 |
| 4/9/03 | Baghdad | 2 | 2 |
| 4/9/03 | Fathlia | 4 | 4 |
| 4/9/03 | Baghdad | 5 | 21 |
| 4/9/03 | Baghdad | 21 | 26 |
| 4/10/03 | Baghdad | 30 | 30 |
| 4/10/03 | Kirkuk | 40 | 40 |

Reason 4:

Events before May 1st, 2003

| Start Date | City | Killings Min. | Killings Max |
|---|---|---:|---:|
| 4/10/03 | Nassiriya | 3 | 3 |
| 4/10/03 | Unknown | 29 | 29 |
| 4/11/03 | Baghdad | 1 | 1 |
| 4/11/03 | Mosul | 2 | 28 |
| 4/11/03 | Baghdad | 3 | 3 |
| 4/11/03 | Nassiriya | 2 | 2 |
| 4/11/03 | Baghdad | 22 | 22 |
| 4/11/03 | Baghdad | 10 | 35 |
| 4/11/03 | Baghdad | 2 | 2 |
| 4/14/03 | Baghdad | 17 | 17 |
| 4/14/03 | Kirkuk | 52 | 52 |
| 4/15/03 | Mosul | 7 | 15 |
| 4/16/03 | Mosul | 3 | 4 |
| 4/18/03 | Baghdad | 1 | 1 |
| 4/19/03 | Baghdad | 3 | 3 |
| 4/19/03 | Baghdad | 3 | 3 |
| 4/20/03 | Tikrit | 0 | 12 |
| 4/20/03 | Kirkuk | 83 | 83 |
| 4/26/03 | Zaafaraniya | 12 | 12 |
| 4/26/03 | Baghdad | 2 | 2 |
| 4/28/03 | Mosul | 0 | 6 |
| 4/28/03 | Fallujah | 13 | 15 |
| 4/30/03 | Fallujah | 2 | 3 |
| Subtotal | | 5504 | 7129 |
| | | | |
| Grand total for all excluded events | | 13876 | 16184 |

**PART 5: An explanation of how the results in this document represent an extensive test, check and confirmation of the findings reported earlier in preprint physics/0506213, which is available at:**

http://xxx.lanl.gov/abs/physics/0506213

This document supercedes preprint physics/0506213 in that it contains:

(1) analysis of additional datasets for other conflicts -- in particular, results for Afghanistan are now included explicitly as an additional figure in the paper,
(2) exhaustive statistical testing in order to check the validity of our empirical findings,
(3) analysis of various extensions of the theoretical model in order to check its general validity,
(4) thorough checking and revision of the datasets for Iraq and Colombia.

The details are as follows:

First, we have improved the Iraq data substantially in several important aspects: we created a new integrated Iraq dataset using multiple sources while avoiding double counting by matching events across sources; we have eliminated non-conflict events (criminal homicides) and composite events that are aggregates of many sub-events; we have added events on coalition casualties in Iraq. These improvements deepen considerably the compatibility between our Iraq data and our Colombia data. Second, we have improved our statistical approach significantly. In the *Appendices*, we present all of our algorithms, tests and statistical significance levels. Interestingly, this switch to more novel and appropriate estimation techniques has not changed our results in any significant way. Third, we have carried out numerous checks of the robustness of our empirical results and have considered many extensions of the mathematical model. Fourth, as in the case of Iraq, we built a new integrated dataset for Afghanistan and successfully obtained a similar power law with an alpha parameter near 2.5 as predicted by our baseline model.

Since our original submission, we have received comments on the following aspects of our work, to which we now respond:

1) Iraq Body Count (IBC) data. Most of the entries in this database are discrete events. But many of them are aggregations of events over stretches of time, sometimes over several weeks. These entries largely break down into two different types.

   a) There are estimates of the number of criminal homicides for a particular month in a particular location. For each time and place IBC takes the number of violent deaths reported by a local morgue and subtracts the average number of pre-war monthly homicides and the number of violent deaths in discrete conflict events included in the IBC database for that time and place. The idea (which we confirmed via direct correspondence with IBC) is that the remainder consists of homicides not picked up by IBC as conflict events and which can be considered "excess deaths" linked to the invasion of Iraq. In other words, these are viewed as part of a wave of criminal homicides unleashed by the coalition invasion.

   For two key reasons, we have purged all of these entries from the version of the dataset we work with in this paper. First, these should be overwhelmingly criminal homicides, not part of the conflict *per se*. In particular, such events are not included in the CERAC Colombia Conflict database so should not be part of the Iraq database for this paper. Second, it is likely that most of the underlying events contributing to the IBC "morgue event" entries are single homicides, or perhaps two or three people killed in a single event. Since our paper is specifically investigating the size distribution of violent events it is a distortion to add these up into single big events.

b) There are other entries that do reflect true conflict activity that are reported as aggregates of underlying events, due to the inability to monitor them on a blow-by-blow basis. Most of these are during the first six weeks of the war, March 20, 2003 to April 30, 2003, when there was a huge aerial bombing campaign. There are two other big composites covering two separate sieges of Falluja in April of 2004 and November of 2004. These entries are measuring true conflict activity so we want them in our database in principle. However, we must be cautious about the impact they can have on our power law estimation. The discussion in the *Appendices* now treats these issues extensively. Here are the main relevant points.

   i) In the paper itself we exclude the first six weeks of the war since during this period the war had more of a conventional than a guerrilla character. So the comparison with Colombia is more appropriate after May 1, 2003 than before. This decision has the consequence of purging almost all the remaining IBC entries that are aggregates of multiple underlying events.

   Nevertheless, in the robustness section of the supplementary materials we recalculate all are results using the full run of data from March 20, 2003. Our results are robust to this extension with the estimated $\alpha$ coming out a bit smaller.

ii) The results in the paper do include the two Falluja entries. However, in the *Appendices* we show that we can remove then with little impact on our results.

In this exercise of checking the robustness of our findings, we also decided to remove a third big event in which many people were killed during a stampede. We have shown that this event can safely be eliminated without significantly changing our results.

In addition, we have introduced further improvements to our Iraq data. The IBC data measure only civilian casualties while the CERAC Colombia data covers both combatants and civilians. To improve their comparability we used data posted on a web site, called iCasualties, to integrate coalition military casualties and contractor personnel casualties into our Iraq data. This required major manual work to ensure that we had no double counting of events. In the course of this work we also discovered a terrorism database, called ITERATE, with a list of Iraq events. So we integrated all three of these. We provide details of on our new dataset in the *Appendices*. We believe that the Iraq and Colombia data is now highly compatible.

Taking a devil's advocate position, one might raise the possibility that the result that the value of the power-law exponent is approximately 2.5, is a coincidence. Although we obviously cannot wait for a re-run of the Iraq or Colombia wars in order to check this, we have addressed this particular concern by building a new dataset for the recent conflict in Afghanistan. This war is believed to have had a similar insurgent character to the Iraq and Colombia wars, and the data is comparable to our Iraq and Colombia data (see *Appendices* for a complete discussion of the database).

Remarkably, we do indeed obtain a power law for Afghanistan with an $\alpha$ of almost exactly 2.5 (see Figure 4 in the paper). In fact, we have now collected data for another four modern conflicts: Indonesia (separatist conflict), Israel-Palestine, Casamance (in Senegal) and Northern Ireland. These all follow power laws with alphas ranging between 1.99 and 3.18, and hence produce an average of around 2.5. Indeed, it is again exactly 2.5 in the case of Indonesia.

In order to investigate further the idea that modern guerilla-like wars are somewhat different from older and/or more conventional wars, we have obtained and analyzed data on three old wars: the civil wars of the US, Spain and Russia. We found no statistical evidence allowing us to distinguish their casualty distribution from another right-skewed, fat-tailed distribution such as a log-normal. We also have analyzed data for the Rwandan genocide and for communal conflict in Indonesia and again we cannot discriminate between different distributions. It therefore seems reasonable to conclude from this extended study that the power-law behavior with an exponent of around 2.5 is indeed a rather reliable property of modern guerilla-like wars, and does not seem to extend well to other conflict settings.

Correspondence with the terrorism data is not exact since this is terrorism data rather than conflict data as used for our Colombia, Iraq and Afghanistan work. Nevertheless, the event is the basic unit of analysis for all four datasets. We think that the fact that terrorism events do not correspond exactly to conflict events adds to the interest of our work since these two different processes turn out to yield very similar size distributions of violent events.

2) Fitting power laws

All our results are now based on a maximum likelihood estimator that takes into account the discreteness of our data. In fact, we did Monte Carlo simulations to compare the properties of this Discrete Maximum Likelihood (DML), with the standard maximum likelihood estimator for continuous data and the OLS estimator. As expected, the DML estimator turns out to be the best of the ones studied and thus is the one we employ in this new paper. It yields both a lower Mean Square Error and lower bias than the other two estimators. The small upward bias of the DML falls rapidly with sample size and it does not depend on $x_{min}$, the minimum value for which the power law holds. Details on the DML and the other two estimators are provided in the *Appendices*. A very encouraging result of this exercise is that even after switching to this estimator, our main findings remain unchanged. In particular the trends reported in the first version of the paper, also turn out to be valid under this change in estimators.

We have also implemented tests for two important hypotheses: first, that the distribution of our data follows a power-law; and second, the hypothesis that it follows a lognormal distribution. We report the statistical significance of these tests. In all three cases (Afghanistan, Colombia and Iraq) we found strong statistical evidence to reject the hypothesis of lognormality but cannot reject the hypothesis of power-law behavior. In addition, we estimate confidence intervals around our point estimates, both for the estimates with the pooled data (Figure 1) and for the picture with the rolling time windows (Figure 2). We have carried out various statistical robustness checks on these, complete with confidence intervals and statistical significance (see the *Appendices*).

3) Mathematical Model

We have carried out extensive analytical and numerical analysis of various generalizations of the model, including extensions to account for multiple armies. We have concluded that the basic result of a steady-state behavior with a power-law near 2.5, is a remarkably robust one. Indeed, it seems to be somewhat of a universal result for coalescence-fragmentation models in which the connectivity between agents is not restricted by geometry (i.e. so-called high-dimensional models) and for a wide class of underlying mechanisms regarding the coalescence-fragmentation process. We are therefore confident about our claims of generality, and regard the result we show as more than just some kind of coincidence. In other words, we truly believe that the fact that the current trend of these guerilla-like wars is to have an exponent around 2.5, is because the current underlying mechanism driving these wars is that of coalescence-fragmentation.

In addition, we supply the following responses to specific criticisms that we have received:

a) "If an attack unit can be a person, an object or a piece of information, then some attack units can be composed completely of objects or information. But clearly, these units pose no threat to anyone without a person to use them in a real attack."

Certainly people will typically be contained in an attack unit (or at least one person) in order for it to be able to attack. Having said this, an attack unit could one day be a person plus a timed explosive device -- then the next day, the person leaves the device by itself (i.e. this attack unit of person-plus-device fragments to give one unit of one person and one unit of a device). The device then detonates itself after say 12 hours. So in the moment of attack, the device has no people in its attack unit. Indeed, in Figure 3, the dark shapes are not meant to represent insurgent fighters themselves, but rather are the potential casualties in an attack by that attack unit. More generally, we stress that our aim is not to model all the precise details of such a conflict. Instead, the mathematical model is designed to deal with average quantities in very general way. In short, it is meant to be a 'minimal model'. Indeed, this is actually a positive aspect of the model, since without needing to specify the detailed contents of the insurgent force, we have obtained good agreement with the observed data merely by invoking a continual process of fragmentation-coalescence. Hence we have managed to identify this process of fragmentation-coalescence as a possible universal driving feature of modern insurgent warfare.

b) "Not every unit with strength S can be decomposed into S units of strength one. For example, a hand gun's strength is not decomposable. Indeed, only if the basic unit of attack is a person do the proposed mechanics of fragmenting and coalescing make any sense. Thus, an attack unit's strength is proportional to the number of people in it, which is only the case if all guerillas have the same information and weapons, and are basically interchangeable."

We agree that a handgun's strength may not be decomposable, although of course the bullets would be. (One bullet can only kill one person, and a handgun without bullets has no attack strength). In addition, explosives are again decomposable into smaller units. Having said this, we again stress that we cannot model all the details of such a conflict. Instead, the mathematical model is designed to deal with average quantities in a 'minimal model' way. Future work will look at the possibility of resupplying the insurgent army, or this army having a fixed initial attack strength which degrades with time. However for the present model, we assume that the total attack strength is maintained around a value $N$ on average, or at least that it changes slowly in time on the scale of the fragmentation-coalescence process. Either of these assumptions will be sufficient for the mathematical results of the model to hold.